\documentclass{aa}
\usepackage{amsmath,array}
\ifx\pdfoutput\undefined
\usepackage{graphicx}
\else
\usepackage[pdftex]{graphicx}
\usepackage{epstopdf}
\fi
\usepackage{txfonts}
\usepackage{natbib}
\bibpunct{(}{)}{;}{a}{}{,}
\usepackage{subfigure}

\begin{document}
\title{Comparison of Methods for modelling Coronal Magnetic Fields}
\author{E. E. Goldstraw\inst{1}\and  A. W. Hood\inst{1} \and P. K. Browning\inst{2} \and P. J. Cargill\inst{1, 3}}
\institute{School of Mathematics and Statistics, University of St Andrews, St Andrews, Fife, KY16 9SS, UK.
\and 
Jodrell Bank Centre for Astrophysics, School of Physics and Astronomy, University of Manchester, Manchester M13 9PL, UK.
\and
Space and Atmospheric Physics, The Blackett Laboratory, Imperial College, London SW7 2BW, UK.
\and
\email{eeg2@st-andrews.ac.uk,  awh@st-andrews.ac.uk}
}
\abstract
{}
{{
Four different approximate approaches used to 
model the stressing of coronal magnetic fields due to an imposed 
photospheric motion are compared with each other and the results from a 
full time-dependent magnetohydrodynamic (MHD) code. 
The assumptions
used for each of the approximate methods are tested by considering large photospheric footpoint displacements.
}}
{{We consider a simple model problem, comparing the full nonlinear magnetohydrodynamic evolution, determined with the Lare2D numerical code, with four approximate approaches. Two of these, magneto-frictional  relaxation and a quasi-1D Grad-Shafranov approach, assume sequences of equilibria, whilst the other two methods, a second-order linearisation of the MHD equations and Reduced MHD, are time-dependent.
}}
{{
The relaxation method
is very accurate compared to full MHD for force-free equilibria for all footpoint displacements but has significant errors when the plasma $\beta_0$ is of order unity. 
The 1D approach gives an extremely accurate description of the equilibria
away from the photospheric boundary layers, and agrees well with Lare2D for all parameter values tested. 
The linearised MHD equations correctly predict the existence of photospheric boundary layers that are present in the full MHD results.
As soon as the footpoint displacement becomes a significant fraction of the loop length, the RMHD method fails to model the sequences of equilibria correctly.
The full numerical solution is interesting in its own right, and care must be taken for low $\beta_0$ plasmas if the
viscosity is too large. 
}}
{}

\maketitle

\section{Introduction}\label{sec:introduction}

{Models of solar flares and coronal heating mechanisms require the build-up and storage of magnetic energy in the coronal magnetic field. This build-up of magnetic energy is frequently modelled by imposing slow photospheric motions that gently stress the coronal field. The common assumption, valid when the driving velocities are very small compared with the coronal Alfv\'en speed, is that the magnetic field will simply pass through a sequence of equilibrium states until the critical conditions, for either an instability or non-equilibrium, are reached and the magnetic energy is subsequently released. 
}

{
Ideally one would like to model this evolution through the full time-dependent non-linear MHD (magnetohydrodynamic) equations. This requires the adoption of a computational approach, but at present, limitations on  resources make the slow evolution over long times difficult to complete. Instead, a variety of approximate approaches that treat the coronal magnetic field in a simplified way have been used. These make different assumptions in order to achieve tractability and it is important to understand how these approaches compare with each other and, especially, how they compare with a full MHD treatment. This has not been carried out before and is the purpose of this paper.}

{To do this, we consider an idealised problem of the shearing of an initially uniform magnetic field in a straightened coronal loop (with the photosphere modelled as two parallel boundaries). Four approximate methods are used, two that consider quasi-static evolution and calculate equilibrium fields and two that consider the time evolution of the field. Note that in the former category, one can calculate a sequence of equilibria in response to footpoint motions, but the intermediate time evolution is lost. The success or otherwise of these approximate models are benchmarked against solutions of the full MHD equations using the Lare computational method
(\cite{arber01}).}

{The first quasi-static methodology considered is the relaxation or magneto-frictional method 
(\cite{yang86,klimchuk92}) which, together with a flux transport model 
 (\cite{mackay06a,mackay06b}), can be used to track the long time evolution of the force-free, coronal magnetic field from days to years. How the field reaches equilibrium is not considered in this approach, but the relaxed state, for the given time evolution of the photospheric magnetic field, is the main goal. This is discussed in Section \ref{subsec:relax}. If there is no equilibrium, for example if a Coronal Mass Ejection (CME) occurs, the relaxation code fails to converge. }

{The second method is based on the well-known idea that two-dimensional equilibria satisfy the Grad
Shafranov equation for the magnetic flux function, $A$; (see Section \ref{subsec:1D}) but, in general, it is difficult to
determine, for specified footpoint displacements, the unknown functional dependencies of the gas pressure 
and the shear component of the magnetic field on $A$. However,
\cite{lothian89} and \cite{browning89} 
used the fact that there is a narrow boundary layer through which the various variables rapidly change from their boundary values to coronal values and that the coronal values only depends on one coordinate. Thus, the two-dimensional approach can be reduced to a one-dimensional problem (in the case when the length of the coronal loop is much larger than the scale of variation of the footpoint motions).
}

{With time-dependent methods, the simplest and most common way to study the evolution is to linearise the MHD equations about a simple initial uniform initial state, as described by \cite{rosner82}. While linearised MHD is straightforward, the possible complexities for this class of problem can be demonstrated by taking the expansion procedure to a higher order. Thus, we can study weakly non-linear effects, due to the non-linear back reaction of the linear solution. The solutions, described in detail in Section \ref{subsec:linear} and the Appendix, will also reveal features that help to justify the use of the one-dimensional solution mentioned above.
}

{Finally, time-dependent non-linear evolution may also be described by the Reduced MHD (RMHD) equations. By eliminating the fast magnetoacoustic waves and utilising the difference in horizontal and parallel length scales, a set of simpler equations can be obtained. RMHD was introduced for laboratory fusion plasma by, for
example, 
\cite{kadomtsev74,strauss76,zank92}
, and used for coronal plasmas by, for example, \cite{scheper99} and \cite{rappazzo10,rappazzo13}. 
A recent review by \cite{oughton17} discusses the validity of the RMHD equations. 
}

{There are a few similar investigations for other situations, for example \cite{pagano13}, who have compared the relaxation method with an MHD simulation for the onset of a CME, \cite{dmitruk05} who compare Reduced MHD with MHD for the case of turbulence and \cite{schrijver06} who test force free extrapolations against a known solution. Examples of footpoint driven simulations include \cite{murawski94,meyer11,meyer12,meyer13}.}

{
Section \ref{sec:methods} describes the simple footpoint shearing experiment, and outlines the details of the four approximate models we examine. Section \ref{sec:results} contains a comparison between these models and benchmarks them against solutions to the full MHD equations. We will find that some methods perform quite well, even when their basic assumptions are not necessarily satisfied. A discussion of the results and possible future benchmarking exercises are presented in Section \ref{sec:conclusions}.}

\section{MHD Equations and Solution Methods}\label{sec:methods}
\subsection{MHD: Basic Equations}
The time evolution of our simple experiment, outlined below, is determined by solving the viscous, ideal MHD equations. The full set of equations are expressed as
\begin{flalign}
&\rho \frac{\partial {\bf v}}{\partial t} + \rho ({\bf v}\cdot \nabla){\bf v} = - \nabla p + {\bf j} \times {\bf B} + \nabla \cdot {\bf S},\label{eq:motion}\\
&\frac{\partial \rho}{\partial t} + \nabla \cdot (\rho {\bf v}) = 0,\label{eq:continuity}\\
&\frac{\partial {\bf B}}{\partial t} = \nabla \times ({\bf v}\times{\bf B}),\label{eq:induction}\\
&\frac{\partial}{\partial t}\left (\frac{p}{\gamma -1}\right ) + {\bf v}\cdot \nabla \left (\frac{p}{\gamma -1}\right ) = - \frac{\gamma p}{\gamma -1} \nabla \cdot {\bf v} + \epsilon_{ij}S_{ij}, \label{eq:energy}
\end{flalign}
together with
\begin{displaymath}
{\bf j} = \frac{\nabla \times {\bf B}}{\mu},\hbox{   and   } \nabla \cdot {\bf B} = 0\; .
\end{displaymath}
${\bf v}$ is the plasma velocity, $\rho$ the mass density, $p$ the gas pressure, ${\bf B}$ the magnetic field and ${\bf j}$ the current density. Gravity is neglected.
The viscous stress tensor is given by
\begin{displaymath}
S_{ij} = 2\rho \nu \left (\epsilon_{ij} - \frac{1}{3}\delta_{ij}\nabla \cdot {\bf v}\right ), 
\end{displaymath}
where $\nu$ is the viscosity and the strain rate is
\begin{displaymath}
\epsilon_{ij} = \frac{1}{2} \left (\frac{\partial v_i}{\partial x_j} + \frac{\partial v_j}{\partial x_i}\right ).
\end{displaymath}
Equations (\ref{eq:motion}) - (\ref{eq:energy}) conserve the total energy, 
$E = \frac{1}{2}\rho v^2 + \frac{B^2}{2\mu} + \frac{p}{\gamma -1}$, so that the dissipation of kinetic energy must go either
into an increase in magnetic energy or an increase in internal energy (i.e. the gas pressure) {defined as $\bf{e=\frac{p}{\gamma-1}}$}. 
The form of the viscous stress tensor does not include the anisotropies introduced by the magnetic field. However {for this experiment}, the main role of the viscosity is to damp out the waves generated by the boundary motions and to allow the
field and plasma to evolve through sequences of equilibrium states so its exact form is not essential. Resistivity is not included as, in general, it will decrease the magnetic energy. \textbf{It has been confirmed that numerical resistivity is negligable as the energy injected at the boundaries equals the energy in the system within $\sim 1\%$.} The aim is to 
follow a sequence of magnetostatic equilibria.

It is normal to express the variables in the MHD equations
in terms of non-dimensional ones and look for dimensionless parameters in the system. Then, it may be possible to use the fact that these parameters are either very large or very small to determine approximate
solutions. Hence, we define a length scale, $R$, a density, $\rho_{0}$, and a magnetic field strength, $B_0$. The dimensionless speed is the Alfv\'en speed, $V_A= B_0/\sqrt{\mu \rho_0}$, and time is expressed in terms
of the Alfv\'en travel time, $t_0= R/V_A$. Hence, we set

\begin{eqnarray}
&&\left ( x, y, z\right ) = R \left (\tilde{x}, \tilde{y}, \tilde{z}\right )\; , \quad t = \frac{R}{V_A} \tilde{t}\; , \quad {\bf B} = B_0 \tilde{{\bf B}}\; ,\nonumber\\
 && p = \frac{B_0^2}{\mu} \tilde{p}\; , \quad {\bf v} = V_A\tilde{{\bf v}}\; , \quad \rho = \rho_0
\tilde{\rho}\; . \label{eq:normalisation}
\end{eqnarray}

Substituting these expressions into equations (\ref{eq:motion}) - (\ref{eq:energy}) and dropping the tildes, the equations remain exactly the same, except that $\mu=1$ and $\nu$ is a non-dimensionless viscosity that is 
the inverse of the Reynolds number. For the values $R = 2\times 10^7$m, $\rho_0 = 1.67 \times 10^{-12}$kg m$^{-3}$ and $B_0 = 10^{-3}$ tesla, the Alfv\'en speed is $V_A = 690$km s$^{-1}$ and the Alfv\'en travel time is
$t_0 = 29$s.

\subsection{Experiment description}\label{sec:expdesc}
Consider a computational box $-l \le x \le l$, and $-L \le y \le L$ and an initial, uniform magnetic field, ${\bf B} = B_0 \hat{y}$, uniform density, $\rho_0$ and uniform pressure, $p_0$. This can be thought of as a coronal
loop of length $2L$ and width $2l$ with a dimensionless plasma $\beta$ equal to $2p/B^2$ and we will use the term \lq loop\rq\  though the results are generic.
In our dimensionless variables, $B_0 = 1$,
$\rho_0 = 1$ and $p_0$ is a constant related to the initial plasma beta, $\beta_0$, {by $\bf{\beta_0=2p_0}$ and initial internal energy by $\bf{e_{0}=\frac{p_0}{\gamma-1}}$}.\\
Now impose a shearing velocity in the $z$ direction at the two photospheric ends ($y= \pm L$). $z$ is chosen to be 
an ignorable coordinate so that the MHD equations will reduce to the appropriate 2.5D form. For the driving motions, we select
\begin{equation}
v_z (x, \pm L, t) = \pm F(t) \sin kx,
\label{eq:shear}
\end{equation}
where $k = \pi/l$ and $v_z(\pm l, y, t) = 0$. The time variation of the shearing velocity is taken as
\begin{equation}
F(t) = \frac{V_0}{2}\left \{\tanh \left (\frac{t- t_1}{\tau_0}\right ) + 1   \right \},
\label{eq:shearvy}
\end{equation}
where $t_1 > \tau_0$ is the switch-on time. We use $t_1 = 6$ and $\tau_0= 2$.
If the parameter $\tau_0$ is small, then $F(t)$ can be approximated by
\begin{equation}
F(t) = \left \{\begin{array}{cc}
                 0\;, & t < t_1\;, \\
                 V_0\; , & t_1 \le t\; .
                 \end{array}\right .
\end{equation}
We can also switch the driving off by using a similar function to ramp down the velocity. 

{This form of the velocity on the boundary will cause the photospheric footpoints to be displaced by a distance $d(x)=D\sin kx$. The maximum footpoint displacement, $D$, can be calculated by integrating the velocity amplitude in time as }
\begin{equation}
D = \int_0^{t} F(t) dt = \frac{V_0 \tau}{2} \left(\log \left \{\cosh \left (\frac{t - t_1}{\tau}\right )\right \} + \frac{t}{\tau}\right ) \approx V_0 (t - t_1)\; ,
\label{eq:displacement}
\end{equation}
for times greater than $t_1$. {Thus, we have three distinct lengths in this problem: the half-length of the loop, $L$; the half-width of the loop, $l$; and the photospheric footpoint displacement, $d(x)$,
from its initial position. In all
cases, we take $L=3$ and $l = 0.3$ so that  $l/L =0.1 \ll 1$. However, we allow $D/L$ to vary from small to large values.}

{
Next, we consider the various speeds in our system. These are: the Alfv\'en speed, $V_A$; sound speed, $c_s=\sqrt{\gamma p_0/\rho_0}$ ($\gamma
=5/3$ is the ratio of specific heats); the speed of the driving motions at the photospheric ends, $V_0$; and a diffusion speed, $V_{visc}=\nu/l$, based on the horizontal lengthscales. 
Typically we take $\nu = 10^{-3}$ so that $V_{visc} \approx 3 \times 10^{-3}$. A smaller value of
$\nu$ could be used but too small a value results in numerical diffusion being more important than the specified value.
In order to pass through sequences of equilibria, we require 
\begin{equation}
V_{visc} \ll V_0 \ll c_s. \label{eq:vineq} 
\end{equation} 
The driving speed is also slow and sub-Alfv\'enic if $V_0 \ll 1$ from Equation (\ref{eq:normalisation}). Accordingly we choose $V_0$ as 0.02. 
Equation (\ref{eq:vineq}) then requires that the pressure is larger than a minimum value of 
 $p_0 \gg 2.4 \times 10^{-4}$. We consider the range
$10^{-3} < p_0 < 1.0$.  Equivalently this can be written in terms of the initial plasma $\beta_0$ as $2\times 10^{-3}<\beta_0 <2.0$ or in terms of the initial internal energy as $\frac{3\times 10^{-3}}{2}<e_{0}<\frac{3}{2}$.}

\subsection{Relaxation}\label{subsec:relax}
Magneto-frictional, relaxation methods solve the induction equation with the velocity given by {the unbalanced Lorentz force} {(see \cite{mackay06a, mackay06b})}. This approach
has had great success in modelling the long term evolution of the global coronal field and in predicting the onset of coronal mass ejections.

To ensure that ${\nabla}{\cdot}{\bf{B}}{=}{0}$, we express the 
magnetic field in terms of a vector magnetic potential, ${\bf A} = ( A_x(x,y), A_y(x,y), A(x,y))$, so that 
\begin{equation}
{\bf{B}} = \nabla \times {\bf{A}} =\left  (\frac{\partial A}{\partial y}, - \frac{\partial A}{\partial x}, \frac{\partial A_y}{\partial x} - \frac{\partial A_x}{\partial y}\right )\; .
\end{equation} 
The equations to be solved are
\begin{flalign}
&{\bf v} = \lambda \frac{{\bf j} \times {\bf B}}{B^2},\label{eq:relaxv}&\\
&\frac{\partial {\bf A}}{\partial t}= {\bf v}\times{\bf B},\label{eq:relaxa}&
\end{flalign}
{where $\lambda=0.3$ is the magneto-frictional constant (see \cite{mackay06a,mackay06b} for details).}

The time evolution is not {physically realistic and is a function of the footpoint displacement} but leads to an end state in which the magnetic field has relaxed to a force-free equilibrium, 
with the imposed $B_z$ from
our shearing displacement. 
{Hence the magnetic energy can be calculated for a given displacement, $D$. 
However, since the velocity is not a realistic quantity the kinetic energy cannot be calculated. }
Once the relaxation process is complete and since the resulting equilibrium is independent of the coordinate $z$, 
the $z$ component of ${\bf A}(x,y)$ is a flux function and the relaxed $z$ component of the magnetic field, $B_z = \partial A_y/\partial x - \partial A_x/\partial y$, 
will be a function of the flux function $A(x,y)$, i.e. $B_z = B_z(A)$. The boundary conditions for the vector potential are
\begin{equation}
A_x(x, \pm L) = \mp B_0  D \sin (\pi x/l)\;  \hbox{   and   } A(x, \pm L) = - B_0 x\; .
\label{eq:potentialbc}
\end{equation}
Without loss of generality, the gauge function is chosen so that $A_y(x, \pm L) = 0$ and, once the field has relaxed, this implies that $A_y(x,y) = 0$.
We select a physical time, $t$, and use Equation (\ref{eq:displacement}) to 
determine the maximum footpoint displacement, $D$. {Note that while solving the Grad-Shafranov equation, Equation (\ref{eq:gradshafranov}) below, for a final force-free equilibrium state, involves only $A$, the evolution
towards such an equilibrium, described by Equation (\ref{eq:relaxa}),} requires calculation of $A_x$ also.
 
\begin{figure*}[ht]
\includegraphics[width=0.95\textwidth]{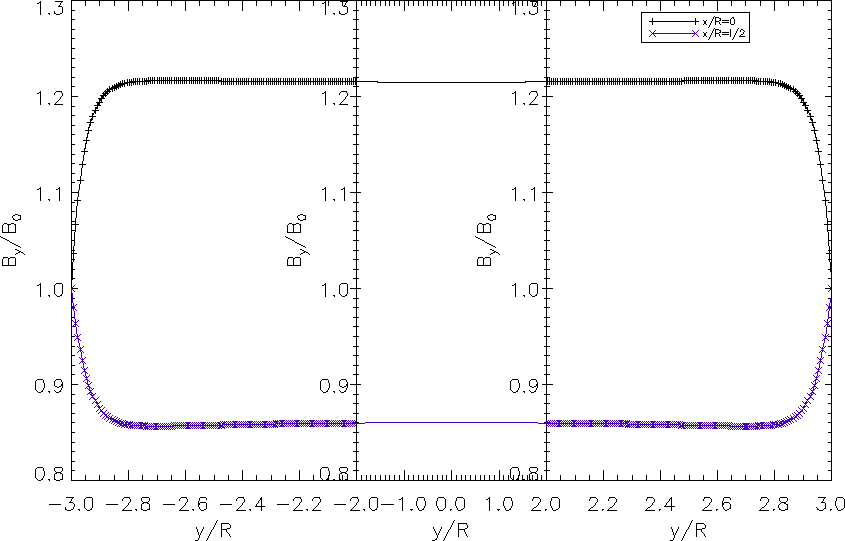}
\caption{{Using the magneto-frictional relaxation method} $B_y/B_0$ is plotted as a function of $y$ for the loop axes $x=0$ (upper) and $x=l/2$ (lower). The horizontal scale is expanded at the two ends to illustrate the resolved boundary layers at $y = \pm L$
and compressed in the middle to demonstrate that there is no variation with $y$ there.}
\label{fig:relax}
\end{figure*}

Given a value of $D$, the magneto-frictional method of \cite{mackay06a,mackay06b}
determines the equilibrium force-free field. For illustration only, we choose $D = 3.0$ (equivalent to $t = 156$) so that $D$ is equal to the half-length $L$.The relaxed state for $B_y$ is shown in Figure \ref{fig:relax}. A more detailed comparison with the other methods is presented in Section \ref{sec:results}.
There are two important points. Firstly, there are sharp boundary layers at the photospheric ends of the field {which have a width of $y/R=0.1$ in this case which equals $l/L$. Different values of this ratio, 0.05, 0.2 and 0.3 have been tested and it is concluded that the
width of these boundary layers is controlled by the width to length ratio, $l/L$. This is also given by the linearised MHD method in Section \ref{subsec:linear}.} 
This shows that $B_y$ rapidly changes from the imposed constant boundary
value of $B_0$ over a short distance that is comparable to the half-width, $l$. Hence, the derivative with respect to $y$ of not just $B_y$ but several variables are large in the boundary layers. 
The width of the boundary layer is \textit{not} dependent on the value of $D/L$ and we use this fact in the next {section when discussing the one dimensional approach}.

 Secondly,
in the middle of the layer, away from the boundaries, $B_y$ is almost independent of $y$ but it does vary with $x$ as $\cos(2kx)$ when $D/L$ is small. 
Thus, although the dominant $y$ component of the field started out uniform, 
when the footpoint displacement is comparable to the length $L$, the variations in $B_y$ are the order of 10\%. Thus, the magneto-frictional method predicts what 
will turn out to be a generic property of relaxed states. 

\subsection{1D Equilibrium}\label{subsec:1D}
When $l/L\ll 1$ a simple estimate of the final equilibrium state is possible, even when the footpoint displacement, $D$, is larger than
the half-length $L$, {$D/L \ge 1$, by solving} the 1D form of the Grad-Shafranov equation. Following the approach of \cite{lothian89, browning89}
and \cite{mellor05}, we can use the fact that the 2D equilibrium can be expressed in terms of the flux function $A(x,y)$ that satisfies the Grad-Shafranov equation
\begin{equation}
\nabla^2 A + \frac{d}{dA}\left (\mu p(A) + \frac{1}{2}B_z^2(A)\right ) = 0.
\label{eq:gradshafranov}
\end{equation}
The pressure is a function of $A$ that is determined by the energy equation and $B_z$ is determined by the shearing introduced by the footpoint displacement. 
For shearing motion\textbf{s} defined in Equations (\ref{eq:shear}) - (\ref{eq:displacement}),
the photospheric footpoint displacement is given by integrating {a fieldline from its initial position, $(x_0,y_0)$, to its final one at $(x,y)$. } 
Hence, it is a function of the flux function and is given by
\begin{eqnarray}
 D(A(x,L)) &&=\int_{y=0}^{y=+L} \left (\frac{B_z(A)}{B_y}\right )_{A=const} dy \nonumber\\
 &&= B_z(A)\int_{y=0}^{y=+L}\left ( \frac{1}{-\partial A/\partial x}\right )_{A=const} dy.
\label{eq:dofA}
\end{eqnarray}
As shown in the above papers and from the magneto-frictional relaxation results, away from the boundaries
we can ignore the boundary layers and assume that the field lines are essentially straight over most of the loop. $l/L \ll 1$ is always assumed.
Away from the boundary layers $A$ is independent
of $y$ and this implies that the integrand is independent of $y$. Therefore, we can determine $B_z(A)$ in terms of the footpoint displacement. Following \cite{mellor05}, we have
\begin{equation}
B_z(A) = -\frac{d(A)}{L}\frac{dA}{dx}.
\label{eq:ByofA}
\end{equation}
For the shearing motion used above, we have on $y=L$, $d(A) = V_0 (t-t_1) \sin kx$, where $k = \pi/l$ and $A(x,L) =  - B_0 x$. Hence, $d(A) = - V_0 (t - t_1)\sin (kx) = - D\sin (kA/B_0)$, where $D = V_0 (t - t_1)$ is the 
maximum footpoint displacement. 

The simple 1D approximation can be modified to include the gas pressure. Conservation of flux and mass between any two fieldlines implies that
\begin{equation}
\frac{B_y}{\rho} = \frac{B_0}{\rho_0}\; , \label{eq:Byrho}
\end{equation}
where $B_0$ and $\rho_0$ are the initial unsheared values.
Next, if the effect of viscous heating is small, the entropy remains constant between any two fieldlines so that
\begin{equation}
\frac{p}{\rho^\gamma} = \frac{p_0}{\rho_0^\gamma}\; .\label{eq:prho}
\end{equation}
Rearranging the last two equations gives the pressure in terms of $B_y$ as
\begin{equation}
p = \frac{p_0}{B_0^\gamma}B_y^\gamma =  \frac{p_0}{B_0^\gamma}\left (-\frac{dA}{dx}\right )^\gamma\; ,
\end{equation}
where $-\partial A/\partial x > 0$.
Hence, the Grad-Shafranov equations reduces to a 1D pressure balance equation of the form
\begin{equation}
\frac{d}{dx}\left (B_y^2 + \left (\frac{D}{L}\right )^2 \sin^2(kA/B_0) B_y^2 + 2 p\right ) = 0\; .
\label{eq:1Deq}
\end{equation}
This implies that the total pressure is constant away from the boundary layers and there is no magnetic tension force.
Computationally, it is easier to express all variables in terms of the flux function, $A$, and solve
\begin{eqnarray}
&&\frac{d^2A}{dx^2}\left (1 + \left (\frac{D}{L}\right )^2 \sin^2 (kA/B_0) +\frac{ \gamma  p_0}{B_0^\gamma} \left (- \frac{dA}{dx}\right )^{\gamma -2} \right ) \nonumber\\
&&= - \frac{k}{2B_0} \left (\frac{D}{L}\right)^2 \sin (2kA/B_0)\left (\frac{dA}{dx}\right )^2\; ,
\label{eq:1D1}
\end{eqnarray}
subject to $A(\pm l) = \mp B_0 l$. The value of the constant total pressure is determined as part of the solution. As shown in Section \ref{sec:results}, this approach provides 
an excellent approximation to the full MHD results for both small and large values of $D/L$.

We can now investigate analytic solutions to Equation (\ref{eq:1D1}) in the extreme cases of small and large $D/L$.
For small shear, $D/L \ll 1$, the solution to Equation (\ref{eq:1D1}) is
\begin{eqnarray}
A &=& - B_0 \left ( x + \left (\frac{D}{L}\right )^2 \frac{\sin (2kx)}{8 k (1 + c_s^2/V_A^2)} \right ) + O\left(\frac{D^4}{L^4}\right )\; ,\\
B_y &=& B_0 \left ( 1 + \left (\frac{D}{L}\right )^2 \frac{\cos (2kx)}{4 (1 + c_s^2/V_A^2)} \right ) + O\left(\frac{D^4}{L^4}\right )\; .
\end{eqnarray}
Hence, the correction to $B_y$ is small (of order $(D/L)^2$). 

For large shear, $D/L \gg 1$, Equation (\ref{eq:1Deq}) is dominated by the middle term, away from $x=0$ and $x=\pm l$. In this case,
\begin{eqnarray}
&&A = -\frac{B_0}{k} \cos^{-1}\left (1 - \frac{2 |x|}{l}\right )\; , \quad B_z = B_0\frac{D}{L} \frac{2}{\pi}\frac{\sin (kA/B_0)}{|\sin(kA/B_0)|} \qquad
\nonumber\\
&& \hbox{and}\qquad B_y =B_0 \frac{2}{\pi}\frac{1}{|\sin(kA/B_0)|}\; .
\end{eqnarray}
$B_z$ has the form of a square wave with value $B_0 (2/\pi) (D/L)$. The minimum value of $B_y$ is $B_0 (2/\pi)$. The variation of the axial field with $x$ is discussed 
along with the other approaches in Section \ref{sec:results}.

\subsection{Time-dependent MHD: 1 Linear and weakly non-linear expansions}\label{subsec:linear}
A simple way to understand some of the properties of the solutions determined above is to linearise the MHD equations about the initial equilibrium state.
We assume that the uniform background magnetic field dominates and we consider small perturbations to this state. 
The expansion is for the case  $B_\perp \ll B_0$, which we expect to be valid when $D/L \ll 1$ and which will be checked a posteriori.
Thus, we  set the form of the expansion as
\begin{eqnarray}
{\bf B} &=& B_0 \hat{\bf y} + B_{1z}(y,t)\sin kx \hat{\bf z} + \left ( B_{2x}(x,y,t) \hat{\bf x}  + B_{2y} (x,y,t)\hat{\bf y}\right )\nonumber \\ &+& \cdots ,\label{eq:Bexp}\\
{\bf v} &=& V_{1z}(y,t)\sin kx \hat{\bf z} + \left ( V_{2x}(x,y,t) \hat{\bf x} + V_{2y}(x,y,t) \hat{\bf y}\right ) + \cdots ,\label{eq:pexp}\\
p &=& p_0 +  p_2(x,y,t) \cdots ,\label{eq:vexp}\\
\rho &=& \rho_0 + \rho_2(x,y,t) \cdots \; , \label{eq:rhoexp}
\end{eqnarray}
where $B_0$, $p_0$ and $\rho_0$ are the constant initial state quantities. 
The subscript \lq 1\rq\ denotes first order terms. Since, {in general, incompressible shearing motions initially only produce Alfv\'en waves,} there is no first order variation in $\rho$
and $p$. The subscript \lq 2\rq\ indicates terms that are second order in magnitude and driven by products of the first order terms, {and are thus weakly non-linear}. The higher order corrections
to the Alfv\'en wave terms will come in at third order.
The expansions break down if the magnitude of the second order terms become as large as the first order terms or if the first order terms are as large 
as the background values. Then, full non-linear MHD must be used.

The MHD equations can now be expanded. To first order, we have the damped Alfv\'en wave equation
\begin{flalign}
&\rho_0 \frac{\partial V_{1z}}{\partial t} = B_0 \frac{\partial B_{1z}}{\partial y} + \rho_0 \nu \nabla^2 V_{1z}, \label{eq:motioneps} &\\
&\frac{\partial B_{1z}}{\partial t} = B_0 \frac{\partial V_{1z}}{\partial y}. \label{eq:inductioneps} &
\end{flalign}

The second order, {weakly non-linear}, equations are
\begin{flalign}
&\rho_0 \frac{\partial v_{2x}}{\partial t} = - \frac{\partial}{\partial x} \left ( p_2 + B_0 B_{2y} + \frac{1}{2}B_{1z}^2 \sin^2 kx\right ) + 
B_0\frac{\partial B_{2x}}{\partial y} \nonumber &\\
 & + \rho_0 \nu \left(\frac{\partial^2 v_{2x}}{\partial x^2}+\frac{\partial^2 v_{2x}}{\partial y^2}
+\frac{1}{3}\frac{\partial}{\partial x}\left ( \frac{\partial  v_{2x}}{\partial x}+ \frac{\partial v_{2y}}{\partial y}\right )\right),\label{eq:motionxeps2}&\\
&\rho_0 \frac{\partial v_{2y}}{\partial t} = - \frac{\partial}{\partial y} \left ( p_2 + \frac{1}{2}B_{1z}^2 \sin^2 kx\right ) \nonumber&\\
 & + \rho_0 \nu \left( \frac{\partial^2 v_{2y}}{\partial x^2}+\frac{\partial^2 v_{2y}}{\partial y^2}+\frac{1}{3}\frac{\partial}{\partial y}\left ( \frac{\partial  v_{2x}}{\partial x}+ \frac{\partial v_{2y}}{\partial y}\right )\right), \label{eq:motionzeps2}&\\
&\frac{\partial \rho_{2}}{\partial t} = - \rho_{0}\left ( \frac{\partial  v_{2x}}{\partial x}+ \frac{\partial v_{2y}}{\partial y}\right ),\label{eq:continuityeps2}&\\
&\frac{\partial B_{2x}}{\partial t} = B_0 \frac{\partial v_{2x}}{\partial y}, \label{eq:inductioneps2x}&\\
&\frac{\partial B_{2y}}{\partial t} = - B_0 \frac{\partial v_{2x}}{\partial x}, \label{eq:inductioneps2y}&\\
&\frac{\partial p_2}{\partial t} = - \gamma p_0 \left (\frac{\partial v_{2x}}{\partial x} + \frac{\partial v_{2y}}{\partial y}\right ) \nonumber &\\
&  + (\gamma -1)\rho_0 \nu \left (k^2 V_{1z}^2 \cos^2 kx 
+\left (\frac{\partial V_{1z}}{\partial y} \right)^2\sin^2 kx \right ).\label{eq:energyeps2}
\end{flalign}
In Equations (\ref{eq:motionxeps2}), (\ref{eq:motionzeps2}) and (\ref{eq:energyeps2}), the linear Alfv\'en wave terms appear as quadratic sources for the second order terms.

\subsubsection{First order solution}
Once the shearing motion starts, an Alfven wave is excited. However, the small viscosity damps this wave and a steady state is reached. To illustrate the ideas for small switch on time, $\bf{t_1}$, 
the solutions to Equations (\ref{eq:motioneps}) and (\ref{eq:inductioneps}) are given by a steady state solution and a Fourier series 
representation of a damped standing Alfv\'en wave. The steady state solution is given by 
\begin{displaymath}
V_{1z }= \left \{
\begin{array}{l l}
      0\; , & t < t_1\; ,\\
      \frac{V_0y}{L}  \sin kx\; , &  t_1 < t\; ,
\end{array}\right.\
\end{displaymath}
and
\begin{equation}
B_{1z} = \left \{
\begin{array}{l l}
      0\; , & t < t_1\; , \\
      B_0\left (\frac{V_0 (t - t_1)}{L} + \frac{\nu k^2L V_0}{2 V_{A}^2}\left (\frac{y^2}{L^2} - 1\right )\right )\sin kx\; , & t_1 < t \; ,
\end{array}\right. \label{eq:BzSteady}
\end{equation}
as can be seen by direct substitution into Equations (\ref{eq:motioneps}) and (\ref{eq:inductioneps}). 
While the solution for $V_{1z}$ 
remains valid for all time, the solution for $B_{1z}$ will be modified once the non-linearities develop. From the maximum
values of our 1D method and Equation (\ref{eq:BzSteady}), we expect the maximum value of $\hbox{max}(B_{z})=B_{max}$, to lie between
\begin{equation}
B_0\frac{2}{\pi}\frac{D}{L} \le B_{max} \le B_0 \frac{D}{L}\; .
\end{equation}
In addition, there are large currents near the photospheric boundaries and numerical resistivity results in field lines slippage (see \cite{bowness13}
and their Equation (24) and Figure 1).

A damped standing wave is required to satisfy the initial conditions, at $t = t_1$, that $V_{1z }= B_{1z} = 0$ for all $y$. The solution for $V_{1z}$ is of the form
\begin{displaymath}
\left ( \frac{V_0y}{L}  + \sum_{n=1}^{\infty} \alpha_n \sin (n\pi y/L) e^{i\omega (t-t_1)}\right ) \sin kx\; ,
\end{displaymath}
where $\omega$ satisfies the appropriate dispersion relation. 
Due to viscosity, $\omega$ is complex and the Fourier series terms decay to zero for large time leaving the steady state solution for $V_{1z}$.
The final steady state for $B_{1z}$ is given in Equation (\ref{eq:BzSteady}). The first term
 depends on the footpoint displacement, $D=V_0 (t - t_1)$. 
The only restrictions
on the maximum speed of the shearing motion of the footpoints are given above {in Section \ref{sec:expdesc}}, namely $V_0$ is greater than the diffusion speed and smaller than the sound and Alfv\'en speeds. 
However, the driving time must be longer than the viscosity damping time
in order to reach a genuine steady state solution. The second term in Equation 
(\ref{eq:BzSteady}) is due to viscosity and is independent of time. In a viscous fluid, if the ends
of the magnetic field are being moved at a speed $V_0$, the central part will lag behind. Hence, $B_{z}$ is smaller in magnitude at $y=0$. This
term will decay after the driving has stopped. What this term does do is produce a gradient in the $y$ direction of 
the magnetic pressure associated with $B_z$ and, although small, it will contribute to a steady flow along the $y$ direction. This is
discussed later.

Using the first order solution, we can calculate the leading order integrated kinetic energy per unit width as a function of time. It is given by
\begin{equation}
\int_{x=-l}^{l} \int_{y=-L}^{L} \frac{1}{2} \rho_0 V_{1z}^2 dy dx = \frac{1}{3} \rho_0 V_0^2 l  L\; .
\label{eq:keintegrated}
\end{equation}
This will be used when interpreting the full MHD, numerical solutions below.
The leading order change to the integrated magnetic energy, however, requires knowledge of second order variables and is discussed below.

\subsubsection{Second order solutions}
Now that the first order steady state solutions are known, the second order equations can be calculated. The terms are complicated although the calculations to generate them are straightforward but tedious.
The details are shown in the Appendix. The basic form of the solutions are given by
\begin{flalign}
&v_{2x}(x,y,t)=(B(y)(t-t_1)+C(y))\sin(2kx)\; ,&\\
&v_{2y}(x,y,t)=(F(y)(t-t_1)+E(y))\cos(2kx)+G(y)\; ,&\\
&B_{2x}(x,y,t)=B_0 \left (B^\prime(y)\frac{(t-t_1)^2}{2} + C^\prime(y) (t-t_1)\right ) \sin (2 k x)\; ,&\\
&B_{2y}(x,y,t)=- 2 k B_0 \left (B(y)\frac{(t-t_1)^2}{2} + C(y) (t-t_1)\right ) \cos(2 k x)\; ,&\\
&\frac{\rho_2(x,y,t)}{\rho_0}= - G^\prime(y) (t-t_1) &\\
& + \left ( [2 k B(y) + D^\prime(y)]\frac{(t-t_1)^2}{2} +(2 k C(y) + E^\prime(y))(t-t_1)\right )\cos(2 k x) \; ,\nonumber& \\
&p_2(x,y,t)= \frac{\gamma p_0}{\rho_0}\rho_2\nonumber&\\
& + (\gamma -1)\rho_0 \nu (t-t_1) \left (k^2 V_{1z}^2 \cos^2 kx 
+\left (\frac{\partial V_{1z}}{\partial y} \right)^2\sin^2 kx \right )\; .&
\end{flalign}
Here $^{\prime}$ denotes a derivative with respect to $y$.
The functions $G(y)$, $B(y)$, $C(y)$, $F(y)$ and $E(y)$ are determined in the Appendix. A key point to note is that $v_{2y}$, when averaged over 
$x$, has a variation in $y$, namely $G(y)$, where
\begin{equation}
G(y) = \frac{\nu k^2 V_0^2 (2 \gamma -1 )}{12 c_s^2} y \left ( \frac{y^2}{L^2} - 1\right )\; .
\label{eq:v2yA}
\end{equation}
Note that for a fixed value of the viscosity $\nu$, this term increases in magnitude if the initial pressure, $p_0$, is reduced. Because of $G(y)$, there is
a change to the density that is independent of $x$, namely 
\begin{equation}
-\rho_0 G^\prime (y) (t-t_1) = \rho_0 \frac{\nu k^2 V_0^2(2\gamma-1)}{12 c_s^2}\left ( 1 - 3 \frac{y^2}{L^2}\right )(t-t_1)\; . 
\label{eq:rhoeps2}
\end{equation}
 Integrating $\rho_2(x,y,t)$ over $x$ and $y$,
we can show that mass is conserved. So the variations of $\rho$ from its uniform initial state are simply a redistribution of the mass through
not only the compression and expansion of the field (variations in $B_y$) but also the flow along fieldlines ($G(y)$). From Equation (\ref{eq:rhoeps2}), the magnitude of this term depends
on the ratio of two lengthscales and two velocities. Defining a diffusion length as $l_d  = \sqrt{\nu (t - t_1)}$, the change in density depends on
\begin{equation}
\pi^2\left (\frac{ l_d}{l}\right )^2 \frac{V_0^2}{c_s^2}.
\end{equation}
As $l_d$ increases with time, $G(y)$ will eventually become important. In addition, it becomes more important for larger $V_0$ and/or lower sound speed, $c_s$.

\subsubsection{Second order solutions: Neglect viscosity}
The expressions for the second order terms are complicated and, for illustration, we simplify them by neglecting viscosity. Setting $\nu = 0$,
\begin{flalign}
&G(y)= C(y) = E(y) = 0\; ,&\\
&B(y) = \frac{\delta}{4k}\left (\frac{\cosh(2ky)}{\cosh(2kL)} - 1\right )\; , &\\
&F(y) = \frac{\delta}{4k}\left ( \tanh(2kL)\frac{y}{L} - \frac{\sinh(2ky)}{\cosh(2kL)}\right )\; , &\\
&\delta = \frac{V_0^2}{L^2}\frac{1}{1 + c_s^2/V_A^2 (1 - \tanh(2kL)/2kL)}\; .&
\end{flalign}
The nature of the boundary layers is clear from the terms, $\cosh(2 k y)/\cosh(2 kL)$ and $\sinh( 2ky)/\cosh(2kL)$, in $B(y)$ and $F(y)$. The width of the boundary layer is controlled by the magnitude
of $2kL$. Hence, the ratio of the half-width to half-length, $l/L$ is important for the size of the boundary layer, as mentioned in Section \ref{subsec:relax}. 
Away from the boundary layers, namely for $2kL \gg 1$, $B(y)\approx -\delta/4k$, $F(y)\approx O(1/2kL)$ and $(1 + c_s^2/V_A^2)\delta \approx (V_0^2/L^2)$ 
and so the second order solutions can be expressed as
\begin{flalign}
&v_{2x}=-\frac{D}{L}\frac{V_0}{4 kL(1+c_s^2/V_A^2)}\sin(2kx)\; ,&\\
&v_{2y}= \frac{D}{L}\frac{V_0}{4 kL(1+c_s^2/V_A^2)} \frac{y}{L}\cos(2kx)\; ,&\\
&B_{2y}=  \frac{D^2}{L^2}\frac{B_0\cos(2 k x)}{4 (1 + c_s^2/V_A^2)}\; , \quad B_{2x}= 0 \; ,\label{eq:B2y} &\\
&\rho_2=  \rho_0 \frac{B_{2y}}{B_0}\; ,\quad p_2= c_s^2\rho_2  = \frac{c_s^2}{V_A^2}B_0 B_{2y}\; .\label{eq:rho2}&
\end{flalign}
Note that Equations (\ref{eq:B2y}) and (\ref{eq:rho2}) agree with the linearised forms of Equations (\ref{eq:Byrho}) and (\ref{eq:prho}) from the 1D equilibrium method. In addition, the second order total pressure, 
$p_2 + B_{1z}^2/2 + B_0 B_{2y}$
is independent of $x$ and equals $(D^2/L^2)(B_0^2/4)$.

From the first order and second order 
magnetic field components, Equations (\ref{eq:BzSteady})
and (\ref{eq:B2y}), the magnitudes of these terms are in powers of $D/L$, making this the appropriate expansion parameter. Hence, these solutions are only strictly 
valid provided $D/L \ll 1$. When viscosity is included,
from Equation (\ref{eq:BzSteady}) the ordering of the terms remains the same provided $\nu < (2 V_A^2/k^2 L V_0)(D/L)$.

The leading order change in the integrated magnetic energy, including the viscosity terms, at second order is given by
\begin{flalign}
&\int_{x=-l}^{l} \int_{y=-L}^{L} \frac{1}{2} B_{1z}^2 dy dx \nonumber,&\\
&= B_0^2 l L \left (\frac{V_0^2 (t - t_1)^2}{L^2} - 
\frac{2}{3}\frac{k^2 \nu V_0^2}{V_A^2} (t - t_1) \right.
  \left.+ \frac{2}{15}\frac{k^4 \nu^2V_0^2 L^2}{V_A^4}\right ),\; \nonumber&\\
& \approx  B_0^2 l L \left (\frac{D^2}{L^2} - 
\frac{2}{3}\frac{D}{L}\frac{k^2 \nu V_0 L}{V_A^2} + \frac{2}{15}\left (\frac{k^2 \nu V_0 L}{V_A^2}\right )^2\right ) \; , &
\label{eq:beintegrated}
\end{flalign}
since the contribution from $B_0 B_{2y}$ integrates to zero. For large $D/L$ or equivalently large time, the magnetic energy is proportional to $(D/L)^2$.

\subsection{Reduced MHD}\label{subsec:rmhd}
Using the Reduced MHD equations and notation quoted in \cite{rappazzo10, rappazzo13} and \cite{oughton17} and assuming that there are no variations in the $z$ direction, 
we can express them as
\begin{flalign}
&\rho_0 \frac{\partial u_x}{\partial t} + \rho_0 u_x\frac{\partial u_x}{\partial x}=-\frac{\partial}{\partial x}\left (p + \frac{b_x^2}{2} + \frac{b_z^2}{2} \right ) + 
B_0\frac{\partial b_x}{\partial y} \nonumber&\\
 & + \rho_0\nu \frac{\partial^2 u_x}{\partial x^2},&\\
&\rho_0 \frac{\partial u_z}{\partial t} + \rho_0 u_x\frac{\partial u_z}{\partial x}= b_x\frac{\partial b_z}{\partial x} + B_0\frac{\partial b_z}{\partial y} + 
\rho_0\nu \frac{\partial^2 u_z}{\partial x^2},&\\
&\frac{\partial b_x}{\partial t} + u_x\frac{\partial b_x}{\partial x}=b_x\frac{\partial u_x}{\partial x} + B_0\frac{\partial u_x}{\partial y},&\\
&\frac{\partial b_z}{\partial t} + u_x\frac{\partial b_z}{\partial x}=b_x\frac{\partial u_z}{\partial x} + B_0\frac{\partial u_z}{\partial y},&\\
&\frac{\partial u_x}{\partial x} = 0, \quad \quad \frac{\partial b_x}{\partial x} = 0.\label{eq:RMHDdiv}&
\end{flalign}
Here we have only included viscosity and, in keeping with the linearised MHD results presented above, we neglect resistivity. The only horizontal derivative included is with respect to $x$.
One consequence of the invariance in the $z$ direction is the prevention of the development of any tearing modes,
which may assist in the creation of short lengths in $z$. ${\bf{B}}_0 = B_0 \hat{\bf{y}}$ is the initial uniform magnetic field
and ${\bf b}$ is the magnetic field created by the boundary motions. \cite{rappazzo10} consider a very similar set-up to this paper.
\cite{oughton17} describe the three main assumptions required for the use of RMHD. These are: (i) the magnetic energy associated with $\bf{B}_0$ is much larger than the magnetic energy associated with $\bf{b}$;
(ii) the derivatives along $\bf{B}_0$ are much smaller than the perpendicular derivatives; and (iii) there are no parallel perturbations so 
that $\bf{B}_0\cdot {\bf b} = 0$ and $\bf{B}_0\cdot {\bf v} = 0$.
Obviously, assumption (i) will fail before the footpoint displacement becomes comparable to the length, $L$, along the initial field. Assumption (ii) 
will hold everywhere, except in the boundary
layers at the two photospheric ends of the field. \cite{scheper99}  have outlined an asymptotic matching procedure to deal with boundary layers in RMHD. They do allow for a
variation in the dominant field component at second order expansion in powers of $l/L$. However, they do not allow for the propagation of the Alfv\'en waves produced by the shearing motions. In
fact, their equations are extremely similar to the magneto-frictional relaxation method described above.
Assumption (iii) will fail before the magnetic pressure variations due to the sheared magnetic field component, $ b_z$,
becomes comparable to $B_0$. Again, this is when the distance the footpoints are moved is the order of $L$. These assumptions are not used
by the methods described above. It is possible that RMHD may be inappropriate because the derivatives in the $z$ direction are in fact 
smaller than the $y$ derivatives. 

From Equation (\ref{eq:RMHDdiv}), the incompressible and solenoidal conditions
simply reduce to $u_x = 0 $ and $b_x = 0$ and not just that they are independent of $x$. The density is assumed to remain constant
and equal to its initial uniform value.
Using Equation (\ref{eq:RMHDdiv}), the above equations simplify to
\begin{flalign}
&0 =-\frac{\partial}{\partial x}\left (p + \frac{b_z^2}{2} \right )\; ,\label{eq:RMHDx}&\\
&\rho_0 \frac{\partial u_z}{\partial t} =  B_0\frac{\partial b_z}{\partial y} + \rho_0 \nu \frac{\partial^2 u_z}{\partial x^2}\; ,\label{eq:RMHDAlfven1}&\\
&\frac{\partial b_z}{\partial t} =  B_0\frac{\partial u_z}{\partial y}\; . \label{eq:RMHDAlfven2}&
\end{flalign}
Equations (\ref{eq:RMHDAlfven1}) and (\ref{eq:RMHDAlfven2}) are similar to Equations (\ref{eq:motioneps}) and (\ref{eq:inductioneps}) in linear MHD 
and describe the propagation of damped Alfv\'en waves.
Once the Alfv\'en waves introduced by the shearing motions have damped, the field passes through
sequences of steady state solutions that are the same as described by the first order linear MHD solutions. 
{In fact the first order linear MHD solutions are exact solutions of the RMHD equations.}

From Equation (\ref{eq:RMHDx}), $p + b_z^2/2$ is constant in the horizontal direction, $x$.{ However, this total pressure is only constant in space and will still} depend on time, as in the 1D method 
presented above. 
Hence, the gas pressure must
balance the $x$ variations in $b_z^2/2$. Such a large gas pressure may not be compatible with a low $\beta_0$ plasma. 
The 1D approach and 
second order solutions, discussed above, include both the gas pressure and the magnetic pressure due to the
modification of $B_y$, namely $B_0 + b_y$. This is a second order change to the uniform magnetic field.
Thus, assumption (iii), that the axial field does not change, must be dropped when the footpoint displacement is sufficiently large.
Instead, it is the total pressure to second order that is constant in $x$, namely
\begin{displaymath}
p + \frac{B_0^2}{2} + B_0 b_y + \frac{b_z^2}{2} = {C}(t)\; .
\end{displaymath}
The constant $C(t)$ must be derived from conservation of flux through the mid-plane. \cite{rappazzo10} do not include $b_y$, where the plasma
forces and evolution depend on the gas pressure gradients and not the current due to variations in $b_{y}$. {Although in some of their cases $b_z$ is very small compared to our values.}

Because $b_y$ is no longer constant, this means that there is compression and expansion. Hence, mass conservation implies that the density
must also change. In a low $\beta_0$ plasma, this is similar to our 1D solution. However,
in the 1D approach, the shear component, $b_z$, is determined by linking the boundary conditions and the footpoint
displacement, through the boundary layers via the flux function $A$.
There is no mention of this in most reduced MHD papers, presumably due to
assumption (ii) that all $y$ derivatives are small compared to the horizontal derivatives. Yet, we know from the relaxation method and,
as will be shown in the full MHD results below, that there can be boundary layers where the $y$ and $x$ derivatives are comparable. 

The solution for $u_z$ is constant in time and has a linear profile between the driving velocity on the lower boundary and the upper boundary. The solution for $b_z$
has two parts to it. The first part is the linear increase in time of the shearing field component while the second part is due to the viscosity term. This is in agreement
with the linearised, first-order solution.

{In summary, care needs to be shown in relating quantities on the boundary to quantities away from
the boundary layers. Many quantities are not the same away from the boundary as they are on the boundary due to the expansion and 
contraction of the magnetic fieldlines.} Hence, it is important when using RMHD, particularly for
simulations in which the boundary footpoints have moved a significant distance in comparison to the length of the field, to check that 
the assumptions in \cite{oughton17} and listed above are indeed satisfied.

\section{Results}\label{sec:results}

Now we briefly summarise each method and clearly distinguish between the many related parameters, $p_0,  \ \beta_0$, $e_{0},\ D,\ t$ before comparing the results.
For full MHD,
 we solve Equations (\ref{eq:motion}) - (\ref{eq:energy}) using the MHD code, Lare2D (see \cite{arber01}), in 2D ($\partial /\partial z = 0$)
for the system described in Section \ref{sec:expdesc} with the driven boundary condition in equations (\ref{eq:shear}) and (\ref{eq:shearvy}).
The width and length of the loop are $l = 0.3$, $L=3$. 
 The photospheric driving speed $V_0 = 0.02$ and the switch on time $t_1 = 6$.
 Viscosity and resistivity are
$\nu = 10^{-3}$
and $\eta=0$. 
{The driving velocity satisfied Equation (\ref{eq:vineq}) so the magnetic
field should pass through a sequence of equilibria. 
This choice means that $V_0$ is slower than the Alfv\'en speed and sound speed, when neglecting slow waves and shocks, but faster than any diffusion speed, as discussed in Section \ref{sec:expdesc}. }

{We have done four simulations each with a different value of $\beta_0$, or equivalently $p_0$ or $e_{0}$. In order to distinguish these related quantities their values are shown in table \ref{tab1}. In the following simulation 1 is referred to as high $\beta_0$ and simulation 3 by low $\beta_0$ unless otherwise stated. This choice has been made for the majority of the results since the other two simulations are qualitatively the same and agree with our understanding in relation to their initial conditions. }
\begin{table}
\caption{The initial internal energy, $e_{0}$ and $\beta_0$ for our four full MHD simulations.}\label{tab1}
\begin{tabular}{m{1.5cm}m{1.5cm}m{1.75cm}}
\hline  Simulation & $\beta_0=2p_0$ & $e_{0}=3/2p_0$ \\
\hline 1&$4/3$ & 1.0  \\
\hline 2& $4/30$& 0.1 \\
\hline 3& $4/300$&0.01\\
\hline 4& $4/3000$&0.001\\
\hline
\end{tabular}

\end{table}

{The maximum displacement, $D$, is related to time, $t$, by Equation (\ref{eq:displacement})
\begin{equation}
D=V_0(t-t_1)\label{eq:D}.
\end{equation}
We choose various times (or equivalently footpoint displacements using Equation (\ref{eq:D})) 
but the times chosen must still be long enough that fast waves
have propagated and equalised the total pressure across the field lines. We present results for cases where the footpoint displacement, $D$, is 
both smaller and larger than $L$, such that $0.29\lesssim D/L\lesssim 2.63$.
The Lare2D results are taken to be the ``exact'' solutions.}

\paragraph{Relaxation:}
\begin{itemize}
\item {As described in Section \ref{subsec:relax}, Equations (\ref{eq:relaxv}) and (\ref{eq:relaxa}) are solved
to evolve the vector potential $\bf{A}$  from an initial state perturbed by the footpoint displacement on the boundaries to a force-free equilibrium.}
\item {Since the actual time evolution of this method is not physical only the magnetic field components for the final state can be compared, hence there are no quantities as functions of time, such as the kinetic energy.}
\item { The perturbation, Equation (\ref{eq:potentialbc}), is determined by the maximum displacement, $D$.}
\end{itemize}
\paragraph{1D Equilibrium Approach:}
\begin{itemize}
\item {The 1D equilibrium approach, described in Section \ref{subsec:1D}, involves solving Equation (\ref{eq:1D1}) for the flux function $A(x,y)$. }
\item {Equation (\ref{eq:1D1}) is determined by the maximum displacement, $D$, and initial pressure, $p_0$. }
\item {This approach gives results for $B_y$, $B_z$, $p$, $j_y$, $j_z$ and $\rho$ as functions of $x$.}
\end{itemize}

\paragraph{Linearisation:}
\begin{itemize}
\item {The first and second order equations and their analytic solution of each variable are described in detail in Section \ref{subsec:linear} and in the Appendix.} 
\item {These expressions are dependent on time, $t$, and the initial pressure, $p_0$.}
\item {\textbf{The solution for each variable consists of the linear and second order terms in order to take into account weakly non-linear effects. These results from linearisation are denoted by ``linear'' in the results section.}}
\end{itemize}
\paragraph{RMHD:}
\begin{itemize}
\item {As discussed {in Section \ref{subsec:rmhd}} RMHD is not applicable to this problem 
but it does agree with the first order terms in linear MHD. }
\item {The first order linear terms are an exact solution to the RMHD equations, Equations (\ref{eq:RMHDAlfven1}) and (\ref{eq:RMHDAlfven2}).}
\end{itemize}

\subsection{Comparison with Lare2D Results}\label{subsec:Lare2Dresults}
We compare all the methods, apart from Reduced MHD, with the full MHD results from Lare2D for the quantities: $B_z$, $B_y$, kinetic and magnetic energy, $\rho$ and $j_y$.  
\subsubsection{Comparison of $B_z$}\label{subsubsec:Bz}
\begin{figure*}[ht]
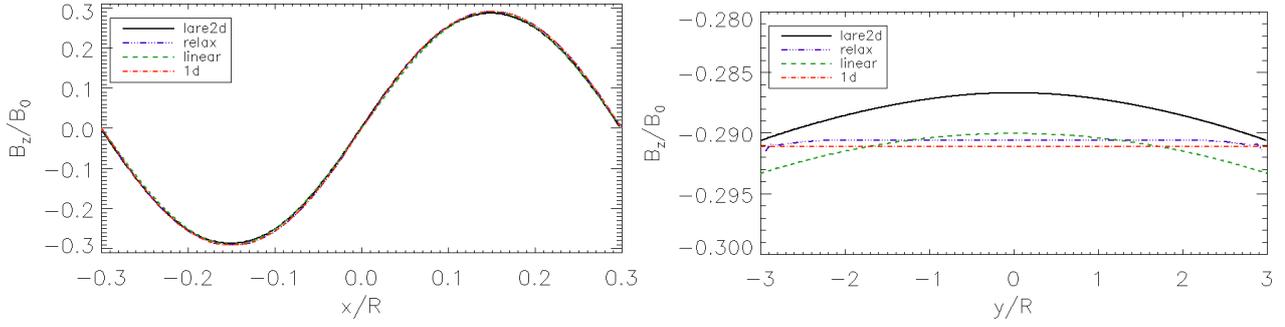

\includegraphics[width=0.45\textwidth]{{{bz_0.9_0.1_x}.png}}
\includegraphics[width=0.45\textwidth]{{{bz_0.9_0.1_y}.png}}
\caption{The sheared magnetic field $B_z$ as a function of $x$ at a midpoint in $y$ (left) and as a function of $y$ for $x=-0.15$ (right) for $\beta_0$ of 4/30 at $t=50$. 
The footpoint displacement is $D/L \approx 0.29$. Solid black curve is for the Lare2D results, triple dot-dashed blue for the relaxation method, dot-dashed green for 1D approximation and  
dashed turquoise for linearised MHD results.}
\label{fig:BzCompare}
\end{figure*}

Firstly we consider the magnetic field component, $B_z$, introduced by the shearing motion. Figure \ref{fig:BzCompare} shows how $B_z$ varies with the horizontal
coordinate, $x$, at the mid-line at $y=0$ (left) and its variation in $y$ at $x=-l/2$ (right) at $t=50$ corresponding to $D/L\approx 0.29$ using Equation (\ref{eq:D}). This is for simulation 2 in table \ref{tab1} which has a reasonably small plasma $\beta_0$ and
the resulting magnetic field will be approximately force-free.
All of the approximations are shown in Figure
\ref{fig:BzCompare}. In fact, the agreement of the $x$ dependence (left part of Figure \ref{fig:BzCompare}) between the methods is remarkably good. This is surprising since the
ratio of $D/L$ is 0.29, which is not particularly small. Hence, one would expect the non-linear terms to be important and the first and second order linear MHD to fail. 
All the methods give good agreement with Lare2D for this value of the plasma $\beta_0$. In the right part of Figure \ref{fig:BzCompare}, the variation with $y$ is shown at $x = -l/2$. As predicted by the linearised MHD expressions
above, there is a slight variation of $B_{z}$ with $y$ that agrees with the Lare2D results. However, the linear results do not include the slight slippage of $B_{z}$ 
at the photospheric boundaries due to the
strong boundary layer currents and so the two curves are slightly displaced. This $y$ variation is not predicted by the 1D and relaxation methods, either 
because they do not use viscosity or it has a different form. 

\begin{figure*}[ht]
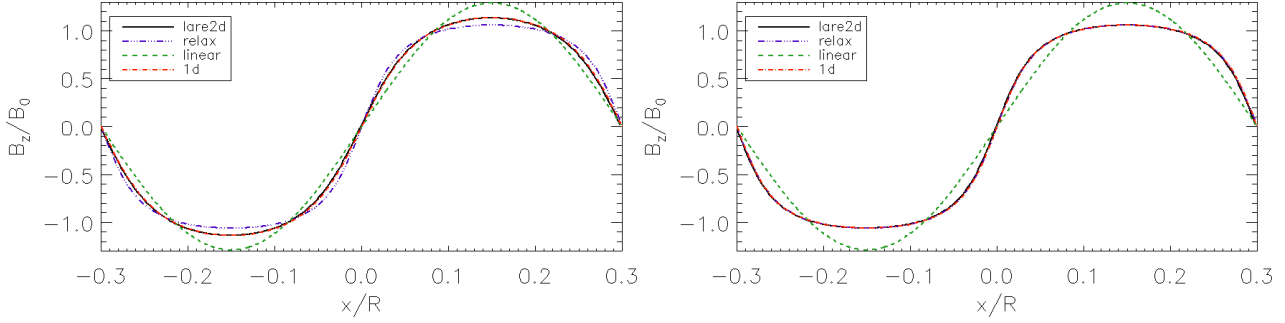

\includegraphics[width=0.45\textwidth]{{{bz_3.9_1.0_x}.png}}
\includegraphics[width=0.45\textwidth]{{{bz_3.9_0.01_x}.png}}
\caption{Plots of $B_z$ against $x$ at the mid-line $y=0$ for each method. The time $t= 200$ and the footpoint displacement is $D\approx 3.9$. 
Left: $\beta_0$ of 4/3 and right: $\beta_0$ of 4/300. }
\label{fig:BzCompare1}
\end{figure*}

When the footpoint displacement is larger than $L$, the shape of the $B_z$ profile changes due to non-linear effects and it takes on an almost square wave structure. This is
shown in Figure~\ref{fig:BzCompare1} for $D/L \approx 3.9/3.0 = 1.3$ ($t = 200$). The large gradients near $x=0$ 
correspond to an enhanced current component, $j_y$, there (shown in figure \ref{fig:currenty} and Section \ref{subsubsec:current}). The left figure is for high $\beta_0$ 
and, for such a large plasma $\beta_0$, the relaxation method has a slight difference {compared to the Lare2D results}. However, this discrepancy is not present in the right figure which is for
low $\beta_0$. In both figures, the linear
approximation is still remarkably good, while the 1D approximation and relaxation sit on top of the Lare2D results. The maximum value of $B_z$ is now about 
unity for both energies and so it is definitely comparable in 
magnitude to the initial background field strength. The RMHD results are not included but they are the same as the linear MHD results.

\subsubsection{Comparison of $B_y$}\label{subsubsec:By}
\begin{figure*}[ht]
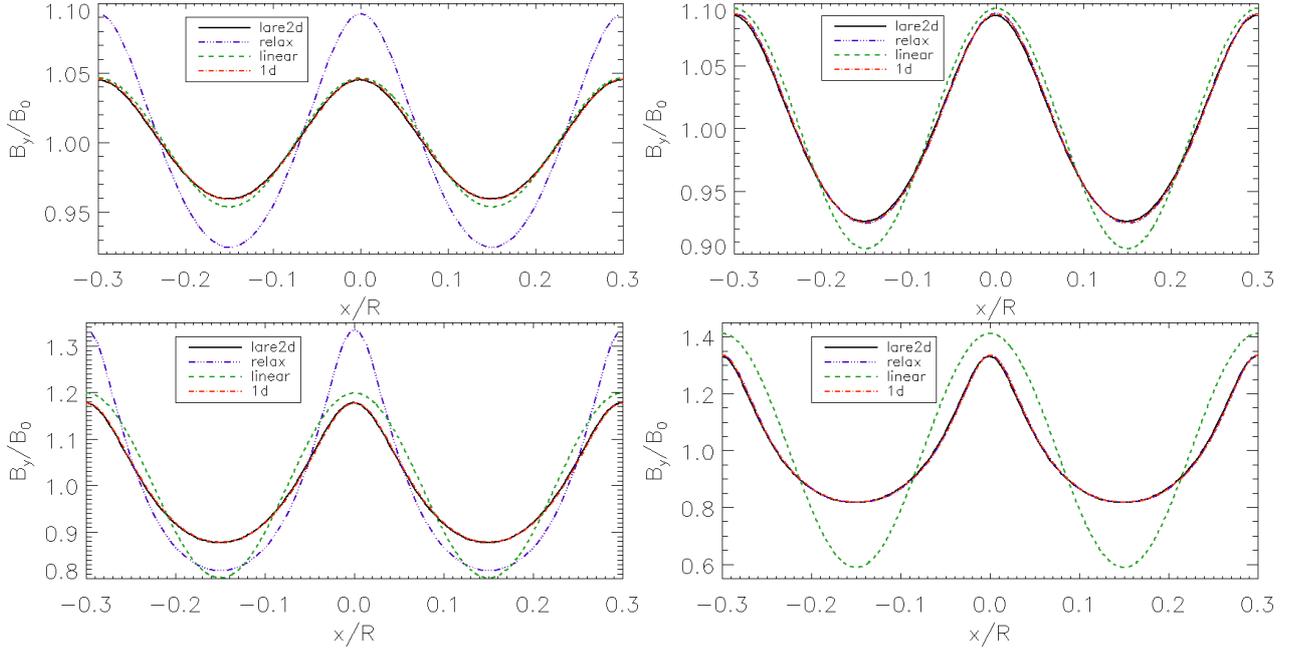

\centering
\includegraphics[width=0.45\textwidth]{{{by_1.9_1.0}.png}}
\includegraphics[width=0.45\textwidth]{{{by_1.9_0.01}.png}}
\includegraphics[width=0.45\textwidth]{{{by_3.9_1.0}.png}}
\includegraphics[width=0.45\textwidth]{{{by_3.9_0.01}.png}}
\caption{Plots of $B_y$ against $x$ in the midpoint in $y$ for $\beta_0=4/3$ (left column) and 4/300 (right column) for 
$D\approx 1.9$ at $t=100$ (top row)
and $D\approx 3.9$ at $t=200$ (bottom row).}
\label{fig:ByCompare}
\end{figure*}

$B_y$ is initially the only magnetic field component.
Figure \ref{fig:ByCompare} shows $B_y$ as a function of $x$ at the mid-line at $y=0$ for $D/L \approx 0.63$ ($t=100$) in the top row 
and $D/L \approx 1.3$ ($t=200$) in the bottom row corresponding to high $\beta_0$ in the left column and low $\beta_0$ in the right column. The other 
parameters are the same as above. 

{For the Lare2D results with small $D/L\approx 0.63$ the maximum value of $B_y$ is about 5\% larger than the initial value for high $\beta_0$ and 10\% for low $\beta_0$, where non-linear effects are becoming important. 
Hence, for footpoint displacements smaller than the loop length the variations in $B_y$
are not too significant. For the case of large $D/L\approx 1.3$ the maximum of $B_y$ is about 20\% larger for high $\beta_0$ and 30\% for low $\beta_0$. It can be concluded that for large values of $D$ any assumption that the horizontal variations in the background field are small is not valid.}

{For the large plasma $\beta_0$ case,
(left column), only the relaxation results are significantly different from the others for both small and large $D/L$, as expected, since this method assumes the field is force-free. 
 Similarly as for $B_z$, in the low $\beta_0$ regime in the right column, the relaxation method agrees with both Lare2D and the 1D approach regardless of the value of the footpoint displacement, $D$.  Interestingly, the
approximation for $B_z$, the shear component, is consistently better than the $B_y$ component, whereas one may expect the same accuracy for both components.}

{The Lare2D and 1D approach both agree with each other extremely well for $4/3000 < \beta_0 < 4/3$ and for $D/L < 2.6$, the largest value
tested.}

{The first and second order linearised MHD agrees reasonably well with Lare2D for small $D/L\approx 0.63$ and high $\beta_0$. For low $\beta_0$ the linear MHD results show a more noticeable discrepancy for small displacement.
For large footpoint displacements, $D/L \approx 1.3$ ($t=200$) in the bottom row, the second order linearised MHD results predict a minimum value of $B_y$ that is too small by
about 10\% for high $\beta_0$ and  25\% for low $\beta_0$ as the non-linear terms become more important. }

{ For Reduced MHD, this component is assumed to remain unchanged during the shearing motion. However, we have shown in the other methods this is not the case and variations become significant after a short time.}
\subsubsection{Comparison of Integrated Energies}\label{subsubsec:kinetic}
\begin{figure*}[!ht]
\centering
\includegraphics[width=0.45\textwidth]{{{eb_1.0}.png}}
\includegraphics[width=0.45\textwidth]{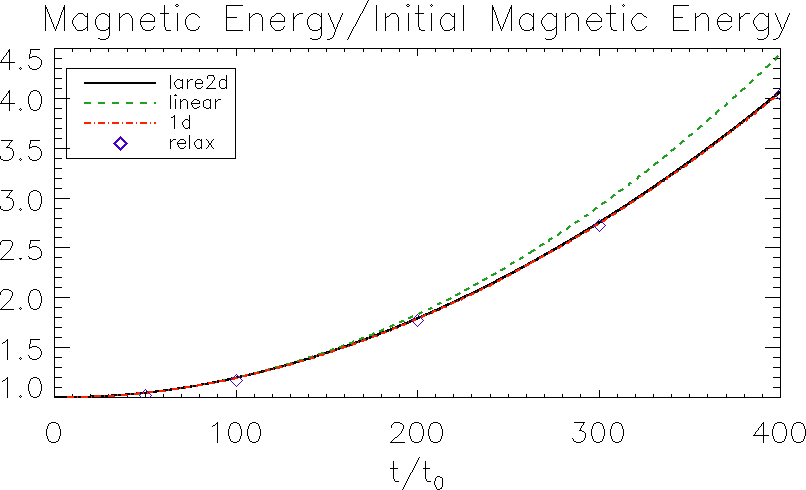}
\caption{{The integrated magnetic energy as a function of time, $t$.  $\beta_0$: left 4/3, right 4/300.}}
\label{fig:Benergy}
\end{figure*}

The integrated magnetic energy is shown in Figure \ref{fig:Benergy} as a function of time for high plasma $\beta_0$ ($\beta_0 = 4/3$) and low $\beta_0$
($\beta_0 = 4/300$). The Poynting flux associated
with the shearing motion results in the magnetic energy increasing nearly quadratically in time for both values of $\beta_0$. 

{The relaxation approach does not directly give quantities as functions of time.  In order to calculate and compare the magnetic energy the magnetic field needs to relax for every value of the displacement. This is limited by resources so the magnetic energy is only calculated for a few values of $D$, shown as symbols on Figure \ref{fig:Benergy}.  These data points agree well with the Lare2D results. As noted for the other quantities there is a marginal discrepancy for high $\beta_0$ which is not present for low $\beta_0$.}
It is interesting to note that the 1D approach correctly matches the results from 
Lare2D for all times, even when the footpoint displacement is
larger than the half-length, $L$, for example, at $t=400$, $D/L \approx 2.6$ using Equation (\ref{eq:D}).
The analytical estimate from the linearised MHD equations,
given in Equation (\ref{eq:beintegrated}),
shows very good agreement up to $t=200$, $D/L\approx 1.3$ when the footpoint displacement is about equal to the loop length and is only in error by 10\% at $t=400$, $D/L\approx 2.6$. 

Thus comparing with Lare2D, we can conclude that the slow magnetic field evolution is correctly modelled by the relaxation and 1D approach
for all times, provided the width to length ratio, $l/L$, is small, and by the linearised MHD method until the footpoint displacement becomes comparable to the loop length, regardless of the size of the plasma $\beta_0$.
{This is notable since once $D\sim L$ one might not have expected the linearisation approach to be valid. } 
\begin{figure*}[!h]
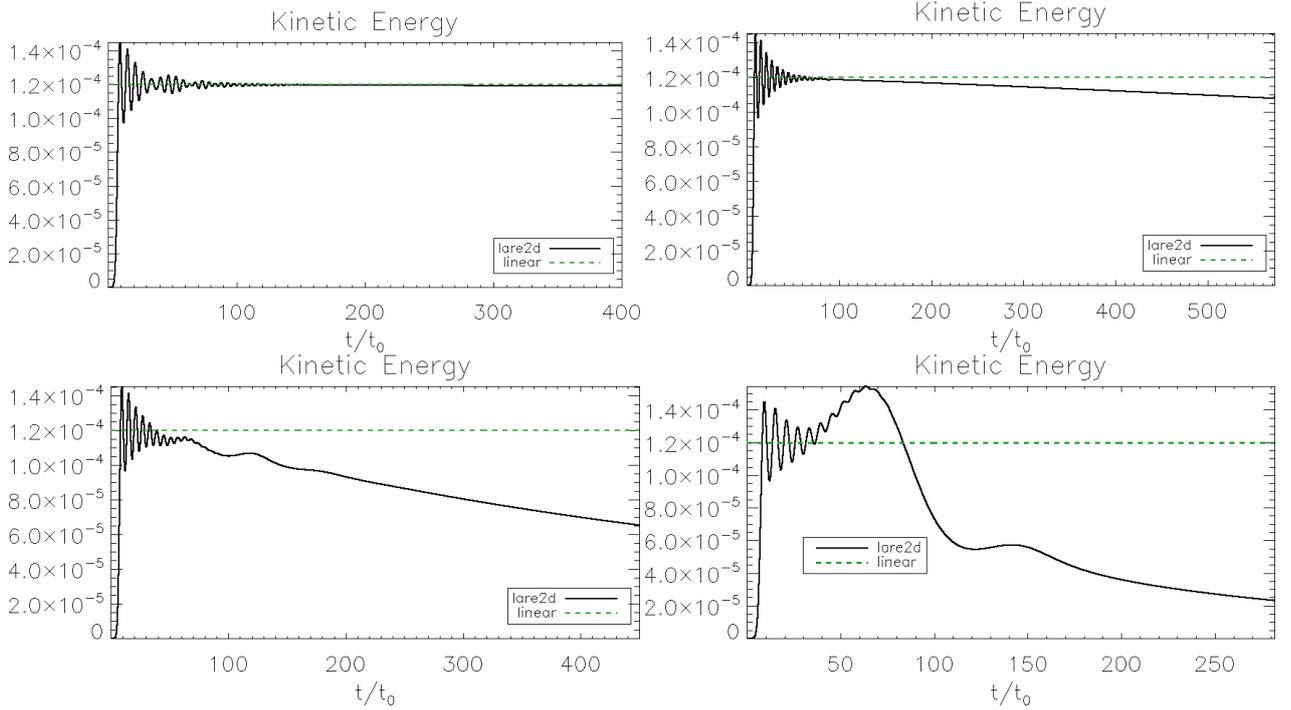

\centering
\includegraphics[width=0.45\textwidth]{{{ke_1.0}.png}}
\includegraphics[width=0.45\textwidth]{{{ke_0.1}.png}}
\includegraphics[width=0.45\textwidth]{{{ke_0.01}.png}}
\includegraphics[width=0.45\textwidth]{{{ke_0.001}.png}}
\caption{The integrated kinetic energy as a function of time, $t$. $\beta_0$: top left 4/3, right 4/30, bottom left 4/300, right 4/3000}
\label{fig:keintegrated}
\end{figure*}

The integrated kinetic energy is shown as a function of time for each of the four different values of the initial plasma $\beta_0$ given in table \ref{tab1} in Figure {\ref{fig:keintegrated}}. The dashed lines are the kinetic energy
estimates given by the first and second order linearised MHD from Equation (\ref{eq:keintegrated}). There are no estimates from either the relaxation method or the 1D approach, as they are assumed to be in equilibrium.
The constant value is only obtained when the Alfv\'en waves, those excited by the switch on the boundary driving velocities, are 
dissipated. Because the driving velocities are slow, the integrated kinetic energy is five orders of magnitude smaller than the magnetic energy. 

What is surprising,
at first sight, is that the Lare2D results only really match the prediction from Equation (\ref{eq:keintegrated}) for an initial high $\beta_0$ plasma. 
As $\beta_0$ is reduced, the departure
from the constant kinetic energy is much more significant. The reason for this departure is due to the flow along the initial magnetic field direction, $v_{y}$ (as shown analytically by the linearized MHD method in Section \ref{subsec:linear}), that is a consequence of 
the magnetic pressure gradient in $y$ due to the $y$ variation in $B_z$ (see Equation (\ref{eq:BzSteady})). The size of the constant flow, $G(y)$, in the second order solution, 
is proportional to $(l_d/l)^2(V_0/c_s)^2$, where the diffusion lengthscale, $l_d$, is defined above and $c_s^2 = \gamma p_0/\rho_0$ is proportional to the initial gas pressure. The viscosity may
be either real or due to numerical dissipation. In both cases, the viscosity damps out both the fast and Alfv\'en waves generated by the switch on of the driving. Once these waves are damped,
the plasma can pass through sequences of equilibria. Although $\nu$ is small, $l_d$ will become large eventually and, so, $p_0$ cannot be too small or else this change in density will occur sooner. 
This steady flow is due to the magnetic pressure gradients introduced by viscosity in the shearing component of the magnetic field, $B_{1z}$. Although the magnitude of this flow is small, it is constant in
time and eventually it will modify the plasma density (see Section \ref{subsubsec:rho} and the second order Equation (\ref{eq:rhoeps2})). In turn, the change in the density will influence the integrated kinetic energy. 

\subsubsection{Comparison with $\rho$}\label{subsubsec:rho}
\begin{figure*}[h]
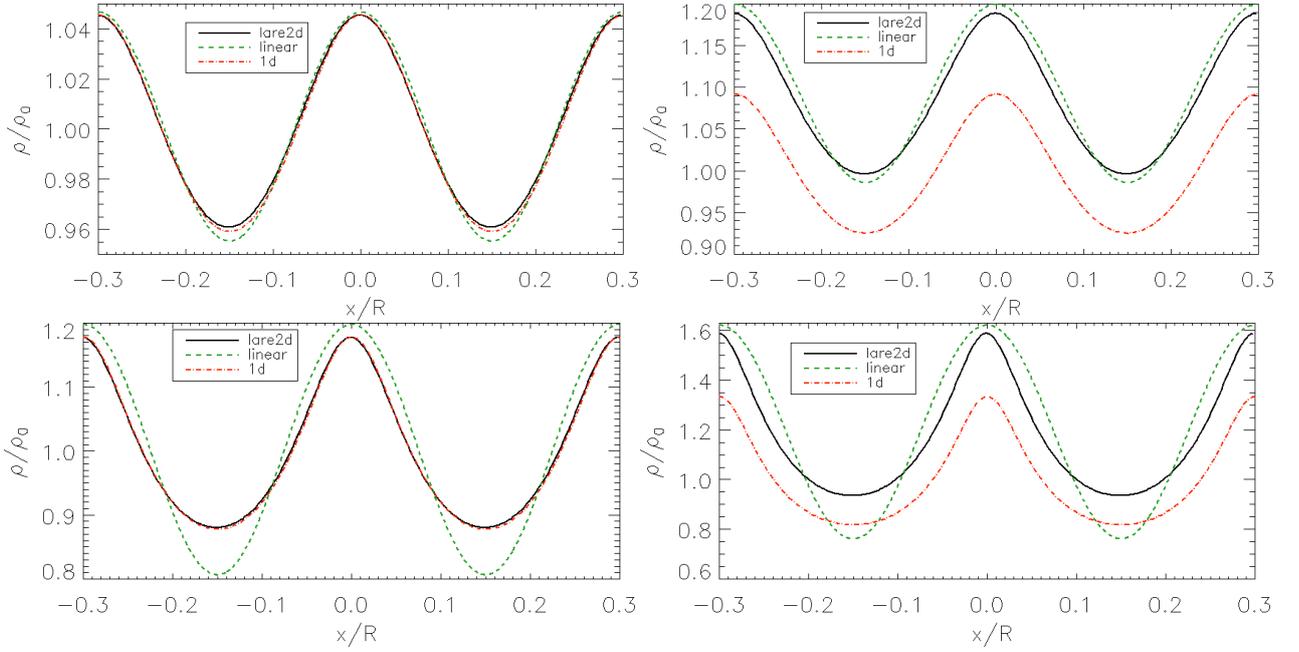

\centering
\includegraphics[width=0.45\textwidth]{{{rho_1.9_1.0}.png}}
\includegraphics[width=0.45\textwidth]{{{rho_1.9_0.01}.png}}
\includegraphics[width=0.45\textwidth]{{{rho_3.9_1.0}.png}}
\includegraphics[width=0.45\textwidth]{{{rho_3.9_0.01}.png}}
\hfill
\caption{Plots of $\rho$ against $x$ at the midpoint in $y$ for $\beta_0$ of 4/3 (left) and 4/300 (right) for $D\approx 1.9$ at time $t= 100$ (top)
and $D\approx 3.9$ for $t=200$ (bottom).}
\label{fig:rhocompare}
\end{figure*}

{The comparison of the plasma density between the Lare2D results, the linearised MHD method and 1D approach is shown in Figure \ref{fig:rhocompare} at the midpoint
in $y$
for high plasma $\beta_0$ (left column), low $\beta_0$ (right column) and footpoint displacement of $D/L \approx 0.63$ (top row) and $D/L\approx 1.3$ (bottom row).
 The relaxation and
RMHD methods are not considered as they do not account for variations in density. 
For high $\beta_0$ and small $D/L$ the agreement
between the three methods is very good. 
The density variations in the $x$ direction are the order of 4\% and all three methods give essentially the same results. 
However, when the plasma $\beta_0$ is small (right column), the density variations
are now between 10\% and 20\% of the initial uniform value, with Lare2D having a general increase in the average value at $y=0$. This is due to the variation in $y$ of $B_z$. These large variations show  that non-linear effects are already becoming important.}
{ For high $\beta_0$ and larger footpoint displacement of $D/L \approx 1.3$ (bottom row) the variations are similar to the 
 low $\beta_0$ case for small $D/L$. This shows that the high $\beta_0$ plasma will eventually evolve in the same way but over a much longer time.
 Once the footpoint displacement has become large the variations in $\rho$ for low $\beta_0$ are nearly 60\% of the initial uniform value of 1.0, thus are very significant}
 
{The 1d approach agrees with Lare2D for high $\beta_0$ for both small and large $D/L$.
 In the case of low $\beta_0$ this method predicts the same variation as full MHD but is displaced slightly which is due to the fact that velocity effects are not included in this approximation.}
  
{The first and second order linearised MHD results agree reasonably well for small displacement for both high and low $\beta_0$.
  In the case of larger $D/L$ the linear results show a difference with the Lare2D results for both large and small
$\beta_0$ as non-linear effects become important.}
\begin{figure*}[ht]
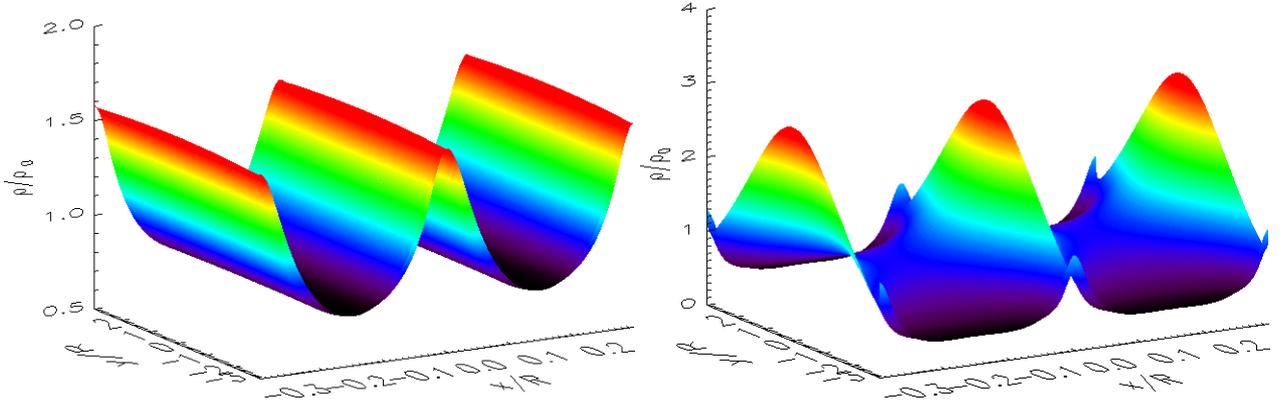

\centering
\includegraphics[width=0.45\textwidth]{{{rhosurf_040_1.0}.png}}
\includegraphics[width=0.45\textwidth]{{{rhosurf_040_0.01}.png}}
\hfill
\caption{Surfaces of density at $t=400$ ($D \approx 7.9$) for $\beta_0$ 4/3 (left) and 4/300 (right).}
\label{fig:rhosurface}
\end{figure*}

{The density dependence on the $y$ coordinate was predicted by the second order solution in Equation (\ref{eq:rhoeps2}). This variation is clearly seen
in the results of Lare2D and this is shown as a 2D surface of $\rho$ in  Figure \ref{fig:rhosurface}, at $t=400$ ($D/L\approx 2.6$), for high $\beta$ (left) and the low $\beta_0$ case (right). The maximum variation in density increases almost linearly in time
and by $t=400$ there is a 15\% difference between the maximum and minimum values at $x=0$. This $y$ variation is not the same as the rapid boundary layer behaviour
seen previously. On the other hand, $\rho$ has almost no $y$ dependence for high $\beta_0$ values (left part of Figure \ref{fig:rhosurface}). This clearly illustrates the large variation in density at $y=0$ as shown in Figure \ref{fig:rhocompare}.
 This is what causes the kinetic energy to decrease as discussed in Section \ref{subsubsec:kinetic}.}
The variations in $\rho$ and $B_y$ will modify the Alfv\'en speed and this can affect the propagation of MHD waves in this plasma.

\subsubsection{Comparison with $j_y$}\label{subsubsec:current}
\begin{figure*}[h]
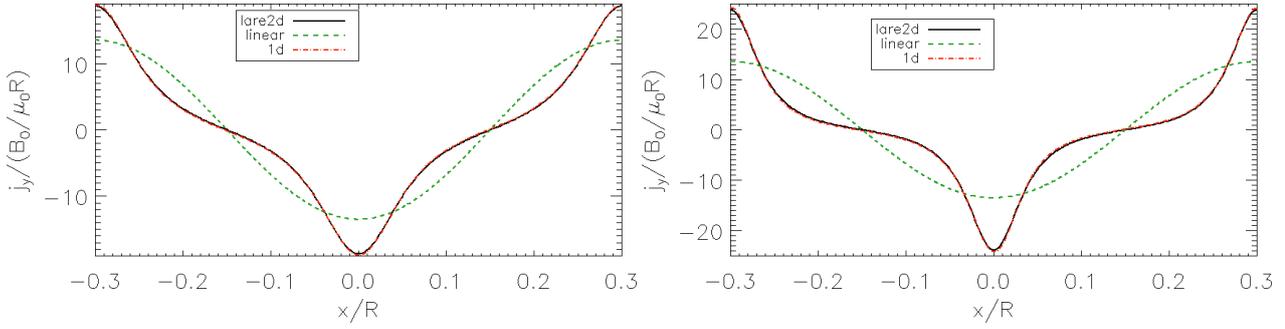

\includegraphics[width=0.45\textwidth]{{{jy_3.9_1.0}.png}}
\includegraphics[width=0.45\textwidth]{{{jy_3.9_0.01}.png}}
\caption{Comparison of $j_y$ against $x$ at the midpoint in $y$ for $\beta_0$ 4/3 (left) and 4/300 (right) for $D \approx 3.9$ ($t = 200$).}
\label{fig:currenty}
\end{figure*}

The current density is an important quantity to determine correctly not only for both force balance but also for ohmic heating, $\eta j^2$. The dominant component of the current density is the $j_{y}$ component
given by $j_{y} = - (\partial B_z/\partial x)$. The results of $j_y$ for Lare2D, 1d approach and linearisation are shown { in Figure \ref{fig:currenty} for $D/L\approx 1.3$ for the case of high $\beta_0$ (left) and low $\beta_0$ (right). The current could be obtained from the relaxation method but has not been done here.}
It is clear that the 1D approach matches the Lare2D results and that the magnitude of the current values exceeds the linear MHD (and also the RMHD)
estimate by almost a factor of 2 (right part of Figure \ref{fig:currenty}) for low $\beta_0$ values. In general, the magnitude of the current increases as $\beta_0$ decreases.
The other component of the current, $j_z = (\partial B_y/\partial x) $ is smaller in magnitude than $j_{y}$ and again both 
Lare2D and the 1D approach agree. 
RMHD does not predict any value for $j_z$.

\section{Conclusions}\label{sec:conclusions}
A simple footpoint shearing experiment has been investigated to test four different methods against full MHD results of Lare2D (\cite{arber01})
and contrasting the other methods with it. This is the first detailed comparison between the different methods, although \cite{pagano13} have compared
the relaxation method with an MHD simulation for the onset of a CME {and \cite{dmitruk05} have compared Reduced MHD and full MHD in the case of turbulence.}

Two methods assume that the magnetic field passes through a sequence of equilibria, namely the magneto-frictional relaxation and 1D methods. The
relaxation method, in the present form, only studies force-free fields and it provides an excellent match to the Lare2D results for $\bf{B}$ for low $\beta_0$, regardless of the footpoint displacement. The inclusion of the gas pressure and
plasma density is possible (see \cite{hesse93}) but has not been done here. 

The second equilibrium method is the 1D approach, which assumes the boundary layers at the photospheric footpoints are narrow and so reduces the
Grad-Shafranov equation to a simple 1D equation for the flux function. 
 Solutions to the resulting equation give outstanding agreement with the Lare2D results for $B_y$, $B_z$, $p$, $\rho$, $j_y$ and $j_z$ for all footpoint displacements and values of $\beta_0$. 
The 1D approach is, of course, derived with this specific experiment in mind. It has
been used for the twisting of coronal loops with cylindrical symmetry (see \cite{lothian89, browning89}). The flux function, is this case, is a function of radius alone.
Unlike the relaxation method, it is not readily extendable to more complex photospheric footpoint displacements but it does do exceptionally well for this particular problem.

{The simplest dynamical approach is to expand the MHD equations in powers of  $D/L$, the ratio of the maximum footpoint displacement to the loop half-length. 
In principal, this should 
only be valid for $D\ll L$. Surprisingly, it has been found that this method provides good agreement for $D/L\lesssim 1$.
One strength of this model is that it can provide useful insight into the system. 
}

Next, we consider Reduced MHD. In general RMHD is identical to the first order linear MHD results, and thus is not capable of reproducing the results from Lare2D. This is primarily because its main assumptions do not hold in this situation. 
While RMHD has the same current component, $j_y$,
as linear MHD, it does not provide any information about, $j_z\approx (\partial B_y/\partial x)$, {since there  can be no change to $B_y$. }
Hence, force balance can only be maintained by balancing the magnetic pressure due to $B_{1z}^2/2$ by  the gas pressure,
instead of through the change to $B_y$.

{There are many possible choices to extend this investigation to more complex systems and to explore the dependencies of this system in more detail.
One question is whether this system is dependent on the form of the internal energy equation.} The gas pressure and density structures produced by the steady shearing motions result in temperature variations. If thermal conduction is included and the boundary conditions keep the temperature
fixed at its initial value, then the temperature will relax towards an isothermal state. The assumption of an isothermal plasma can be included in the 1D method very easily by setting $\gamma = 1$. There is still a variation 
in pressure and density. So the inhomogeneous nature of the resulting plasma is not dependent on the exact form of the internal energy equation.

{Further work on the validity of Reduced MHD is required.
For this experiment, we can neglect the variations parallel to the initial field whenever the horizontal lengthscales are much shorter than the parallel ones. 
However, whether one can use RMHD or not depends on both the final footpoint location, the total displacement and how the field lines got there.  
On the one hand, a simple shear followed by the opposite shear brings
the footpoints back to their initial locations but the field will remain potential. On the other hand, a complete rotation also brings the footpoints to 
their initial locations but
this time the field is not potential. It is how the field gets to the final location and the total Poynting flux that is injected into the corona that is important.}

{The message from this work is that one needs to take care with simply implementing a method without thinking whether the assumptions are valid or not. The four approximate methods have been used for a particularly simple shearing experiment. For example, the simple 1d method is inappropriate for more complex and realistic photospheric footpoint motions. However, the magneto-frictional, relaxation method is still applicable provided the displacement of the footpoints is small from the previous equilibrium state. Hence, a simple rotation of the footpoint through 360 degrees can be achieved by splitting the rotation up into smaller angles and relaxing before taking the next small rotation. For small angular motions, the relaxation method will quickly reach the nearby equilibrium state. This is then repeated until the complete revolution is achieved. See \cite{meyer11, meyer12,meyer13} for the application of the relaxation approach to the velocities derived from the magnetic carpet. The linearization of the MHD equations can always be undertaken but the derivation of an analytical expression for the linear solution with more complex boundary conditions is not certain. Without an expression for the linear steady state, it will be difficult to determine the modifications to the density and main axial field in response to the non-linear driving by the linear steady state. Reduced MHD can certainly be applied to more complex photospheric motions but we would expect that the quadratic terms, due to the linear terms, will invalidate some of the main assumptions stated in \cite{oughton17}. Solving the full MHD equations remains the preferred approach, provided sufficient computing resources are available to generate the long time evolution of the magnetic field.}

\begin{acknowledgements}
The authors thank the referee for extremely useful comments.
AWH and EEG thank Duncan Mackay for useful discussions on the magneto-frictional method.
AWH acknowledges the financial support of STFC through the Consolidated grant,  ST/N000609/1, to the University of St Andrews and EEG acknowledges the STFC studentship,
ST/I505999/1. This work used the DIRAC 1, UKMHD Consortium machine at the
University of St Andrews and the DiRAC Data Centric system at Durham University,
operated by the Institute for Computational Cosmology on behalf of the
STFC DiRAC HPC Facility (www.dirac.ac.uk). This equipment was funded by a BIS National
E-infrastructure capital grant ST/K00042X/1, STFC capital grant ST/K00087X/1,
DiRAC Operations grant ST/K003267/1 and Durham University. DiRAC is part of the
National E-Infrastructure.
\end{acknowledgements}
\begin{appendix}
\section{Second order solutions}\label{sec:app}
Now we can solve the second order Equations (\ref{eq:motionxeps2}) - (\ref{eq:energyeps2}) for $v_{2x}$ and $v_{2y}$. Then we can determine the other variables.
We will include  the viscous heating and dissipation terms.

Taking the time derivative of Equations (\ref{eq:motionxeps2}) and (\ref{eq:motionzeps2}) and using Equations (\ref{eq:inductioneps2y}) and (\ref{eq:energyeps2}), we have
\begin{flalign}
\frac{\partial^2 v_{2x}}{\partial t^2}=&-\frac{\partial}{\partial x}\left( - c_s^2 \left (\frac{\partial v_{2x}}{\partial x} + \frac{\partial v_{2y}}{\partial y}\right )\right.\nonumber&\\
 &\left.+(\gamma-1)\nu \left(\frac{k^2V_0^2y^2}{L^2}\cos^2(kx)+\frac{V_0^2}{L^2}\sin^2(kx)\right)\right )\nonumber &\\
&-\frac{\partial}{\partial x}\left (\frac{V_A^2}{2}\left(\frac{2V_0^2\tau}{L^2}+\frac{\nu k^2V^2_0 }{V_A^2}\left(\frac{y^2}{L^2}-1\right)\right) \sin^2(kx)\right)\nonumber&\\
 & +V_A^2\left(\frac{\partial^2 v_{2x}}{\partial x^2}+\frac{\partial^2v_{2x}}{\partial y^2}\right)\nonumber &\\
& +\nu \frac{\partial}{\partial t}\left(\frac{4}{3}\frac{\partial ^2v_{2x}}{\partial x^2}+\frac{\partial ^2v_{2x}}{\partial y^2}+\frac{1}{3}\frac{\partial^2v_{2y}}{\partial x\partial y}\right)\; ,&
 \label{eq:fastvxeps2}
\end{flalign}
and
\begin{flalign}
\frac{\partial^2 v_{2y}}{\partial t^2} =& -\frac{\partial}{\partial y}\left(- c_s^2 \left (\frac{\partial v_{2x}}{\partial x} + \frac{\partial v_{2y}}{\partial y}\right )\right. \nonumber&\\
 & \left.+(\gamma-1)\nu \left(\frac{k^2V_0^2y^2}{L^2}\cos^2(kx)+\frac{V_0^2}{L^2}\sin^2(kx)\right)\right )\nonumber &\\
& -\frac{\partial}{\partial y}\left (\frac{V_A^2}{2}\left(\frac{2V_0^2\tau}{L^2}+\frac{\nu k^2V^2_0 }{V_A^2}\left(\frac{y^2}{L^2}-1\right)\right) \sin^2(kx)\right) \nonumber &\\
&+\nu \frac{\partial}{\partial t}\left(\frac{\partial^2v_{2y}}{\partial x^2}+\frac{4}{3}\frac{\partial ^2v_{2y}}{\partial y^2}+\frac{1}{3}\frac{\partial^2v_{2x}}{\partial y\partial x}\right)\; ,&
\label{eq:fastvzeps2}
\end{flalign}
where $\tau = t- t_1$.
These two equations can be solved by taking
\begin{flalign}
&v_{2x}(x,y,t)=(B(y)\tau+C(y))\sin(2kx),\label{eq:v2xdriven}&\\
&v_{2y}(x,y,t)=(F(y)\tau+E(y))\cos(2kx)+G(y),&\\
&B(y) =  \frac{\delta}{4k}\left (\frac{\cosh(2ky)}{\cosh(2kL)} - 1\right )\; , &\\
&G(y)=\nu \frac{(2 \gamma -1)k^2 L^2}{12c_s^2}\left (\frac{V_0^2}{L^2}\right ) y \left (1 - \frac{y^2}{L^2}\right )\; ,&\\ 
&C(y)=\frac{\nu}{2kV_A^2}\left(\alpha+\frac{\delta}{2}\right)\left(\frac{\cosh(2ky)}{\cosh(2kL)}\left(2k^2L^2+1\right)-\left(2k^2y^2+1\right)\right)\nonumber &\\
&+ \frac{\nu}{4kV_A^2}\left(2c_s^2\kappa-\delta+\frac{2}{3}\alpha\right)\left(\frac{\cosh(2ky)}{\cosh(2kL)}-1\right)\; ,&\\
&F(y) = \frac{\delta}{4k}\left ( \tanh(2kL)\frac{y}{L} - \frac{\sinh(2ky)}{\cosh(2kL)}\right )\; , &\\
&E(y)=\frac{2\nu k^2}{3}\left(\left(\alpha+\frac{\delta}{2}\right)\left(\frac{1}{V_A^2}+\frac{1}{c_s^2}\right)+\frac{V_0^2\left(\gamma-\frac{3}{2}\right)}{4L^2c_s^2}\right)y^3
\nonumber &\\
&+\nu \left(\frac{ k^2V_0^2}{4c_s^2}+\left(\frac{4\alpha}{3V_A^2}-\frac{V_0^2(\gamma-1)}{2L^2c_s^2}\right)+\kappa\left(\frac{c_s^2}{V_A^2}+1\right)\right)y\nonumber&\\
&-\nu \left(L^2 \left(\alpha+\frac{\delta}{2}\right)k^2+\frac{c_s^2\kappa}{2}+\frac{2 \alpha}{3}\right)\frac{\sinh(2ky)}{V_A^2k\cosh(2kL)}\; ,&
\end{flalign}
where $\alpha,\ \delta$ and  $\kappa$ are constants, chosen to satisfy the boundary conditions, namely 
\begin{eqnarray}
\alpha&=&\frac{\delta}{4kL}\tanh(2kL)-\frac{\delta}{2}\; ,\\
\delta&=&\frac{V_0^2/L^2}{1 + (c_s^2/V_A^2)\left ( 1 - \tanh(2kL)/2kL\right )}\; ,\\
\kappa&=& \frac{( V_1 + V_2)}{c_s^2L\left(kL\left(c_s^2+V_A^2\right)-\frac{1}{2}c_s^2\tanh(2kL)\right)}\; ,
\end{eqnarray}
and
\begin{eqnarray}
V_1 &=& c_s^2L\left(k^2\left(\alpha+\frac{\delta}{2}\right)L^2+\frac{2}{3}\alpha\right)\tanh(2kL)\; ,\\
V_2 &=&-\frac{2}{3}k\left(\left(c_s^2+V_A^2\right)\left(\alpha+\frac{\delta}{2}\right)k^2L^4+\left(\frac{V_0^2\gamma V_A^2k^2}{4}+2\alpha c_s^2\right)L^2\right. \nonumber \\
& &\left.-\frac{3V_0^2V_A^2(\gamma-1)}{4}\right)\; .
\end{eqnarray}
From the expressions for $v_{2x}$ and $v_{2y}$, we can calculate the other variables as
\begin{flalign}
&B_{2x}(x,y,t)=B_0 \left (B^\prime(y)\frac{\tau^2}{2} + C^\prime(y) \tau\right ) \sin (2 k x)\; ,&\\
&B_{2y}(x,y,t)=- 2 k B_0 \left (B(y)\frac{\tau^2}{2} + C(y) \tau)\right ) \cos(2 k x)\; , &\\
& \frac{\rho_2(x,y,t)}{\rho_0}= - G^\prime(y) \tau \nonumber&\\
&+ \left ( [2 k B(y) + F^\prime(y)]\frac{\tau^2}{2} +(2 k C(y) + E^\prime(y))\tau\right )\cos(2 k x) \; ,&\\
&p_2(x,y,t)= \frac{\gamma p_0}{\rho_0}\rho_2\nonumber&\\
& + (\gamma -1)\rho_0 \nu \tau \left (k^2 V_{1z}^2 \cos^2 kx 
+\left (\frac{\partial V_{1z}}{\partial y} \right)^2\sin^2 kx \right )\; .&
\end{flalign}
Notice that $B_{2y}$ and $B_{2x}$ also have boundary layers and that
\begin{equation}
B_{2x} \approx 0\; , \quad B_{2y} \approx B_0 \left(\frac{\delta \tau^2}{4}+\frac{\nu c_s^2\kappa \tau}{V_A^2}\right)2\left (\frac{1}{2} - \sin^2(kx)\right )\; ,
\label{eq:Bzeps2approx}
\end{equation}
in the central part of the field away from the boundary layers.
From Equation (\ref{eq:v2xdriven}), $v_{2x}$ does remain small but it is essential in allowing the axial field to adjust value.
Calculating the magnetic pressure to second order we find that
\begin{eqnarray}
\frac{B_{1z}^2 + (B_0 + B_{2y})^2}{2} &=& \frac{B_{1z}^2}{2} + \frac{B_0^2}{2} + {B_0 B_{2y}}, \nonumber \\
&=& \frac{B_0^2}{2 } \left ( 1 + \frac{\delta \tau^2}{2}+\frac{2\nu c_s^2 \kappa \tau}{V_A^2}+\left[\tau^2\left (\frac{V_0^2}{L^2}-\delta \right) \right.\right. \nonumber\\
&+&\left.\left. \nu \tau\left(\frac{ k^2V_0^2}{V_A^2}\left(\frac{y^2}{L^2}-1\right)-\frac{4c_s^2}{V_A^2}\kappa\right)\right]\sin^2kx\right ).
\label{eq:Bpressure2}
\end{eqnarray}
The magnetic pressure is growing quadratically in time and is dependent on $x$ and $\delta$. The neglected term is the square of the viscous part of $B_{1z}$.
Including the second order gas pressure gives 
\begin{flalign}
p_2+\frac{B_{1z}^2}{2} + \frac{B_0^2}{2} + {B_0 B_{2y}} 
=&\frac{B_0^2}{2}+\frac{B_0^2V_0^2\tau^2}{4L^2}\nonumber\\
&+\frac{\rho_0\nu \left(k^2L^2(\gamma-2)+3(\gamma-1)\right)}{6}\frac{V_0^2}{L^2}\tau.
\label{eq:totalpressure2}
\end{flalign}
 This removes the dependence on $x$ and thus it can be concluded that the \textit{total}
pressure is independent of $x$ but is increasing in time.
We can also calculate the first and second order  current
\begin{eqnarray}
j_{1x}&=&\frac{\partial B_{1z}}{\partial y}\sin(kx)
={B_0}\frac{\nu k^2V_0}{V_A^2L}y\sin(kx), \; \\ 
j_{1y}&=&-{k}B_{1z}\cos(kx),\nonumber\\
&=&-{kB_0}\left(\frac{V_0 \tau}{L}+\frac{\nu k^2LV_0}{2V_A^2}\left(\frac{y^2}{L^2}-1\right)\right) \cos(kx)\: ,\\ 
j_{2z}&=&-2k B_{2y}\sin(2kx)=-2k B_0 \left(\frac{\delta \tau^2}{4}+\frac{\nu c_s^2\kappa \tau}{V_A^2}\right)\sin(2kx)\; .
\end{eqnarray}

\end{appendix}
\bibliographystyle{aa}
\bibliography{RMHD}

\end{document}